\documentstyle[psfig,epsf]{mn}

\begin{document}
\title{Linear and nonlinear contributions to pairwise peculiar velocities}
\author[R. K. Sheth et al.]
{Ravi K. Sheth$^1$, Lam Hui$^{2,3}$, Antonaldo Diaferio$^4$ \& 
Rom\'an Scoccimarro$^2$ \\
$^1$ NASA/Fermilab Astrophysics Group, MS 209, Batavia, IL 60510-0500\\
$^2$ Institute for Advanced Study, School of Natural Sciences, Einstein Drive, Princeton, NJ 08540\\
$^3$ Department of Physics, Columbia University, 538 West 120th Street, 
New York, NY 10027\\
$^4$ Dipartimento di Fisica Generale ``Amedeo Avogadro'', 
Universit\'a di Torino, Italy\\
\smallskip
Email: sheth@fnal.gov, lhui@ias.edu, scoccima@ias.edu, diaferio@ph.unito.it}
\date{Submitted to MNRAS 2000 August 15}

\maketitle

\begin{abstract}
We write the correlation function of dark matter particles, $\xi(r)$, 
as the sum of two terms---one which accounts for nonlinear evolution, 
and dominates on small scales, and another which is essentially the 
term from linear theory, and dominates on large scales.  We use models 
of the number and spatial distribution of haloes and halo density profiles 
to describe the nonlinear term and its evolution.  The result provides a 
good description of the evolution of $\xi(r)$ in simulations.  We then 
use this decomposition to provide simple and accurate models of how the 
single particle velocity dispersion evolves with time, and how the first 
and second moments of the pairwise velocity distribution depend on scale.  
The key idea is to use the simple physics of linear theory 
on large scales, the simple physics of the virial theorem on small 
scales, and our model for the correlation function to tell us how to 
weight the two types of contributions (linear and nonlinear) to the 
pairwise velocity statistics.  When incorporated into the streaming model, 
our results will allow a simple accurate description of redshift-space 
distortions over the entire range of linear to highly nonlinear regimes.  
\end{abstract} 


\section{Introduction}\label{intro}
Strong constraints on models of large scale structure follow from combining 
statistics of the density field with statistics of the velocity field.  
In this paper, we show how the number of particle pairs depends on pair 
separation, and use this to compute the mean and mean square pairwise 
velocity.  That is, we show how the correlation function of the density 
field and the distribution of pairwise velocities can all be computed 
from the same model.

The key to being able to do this is a simple model of how and why the 
correlation function $\xi(r,a)$ evolves with time.  The evolution of 
$\xi$ was first accurately modelled by Hamilton et al. (1991).  
Following Peebles (1980), they showed that knowledge of how the 
correlation function evolved allowed them to describe how the mean 
streaming velocity $v_{12}(r,a)$ of particle pairs evolved as well 
(also see Nityananda \& Padmanabhan 1994).  Peebles suggests that the 
second moment of the pairwise distribution, $\sigma_{12}(r,a)$ depends on 
(an integral over) the three-point correlation function $\zeta(r,a)$; 
because there is no general description of $\zeta$ from linear to 
non-linear scales (but see Scoccimarro \& Frieman 1998 for scale-free 
initial conditions), there is, at present, no simple description of how 
the pairwise velocity dispersion depends on scale (see, e.g., 
Mo, Jing \& B\"orner 1997; Jing, Mo \& B\"orner 1998).  

In this paper, we will follow a different approach than the one laid 
down by Hamilton et al.  The logic behind our approach follows from 
the models first discussed by Neyman \& Scott (1959), and references 
therein.  
In these models, all particles are assumed to be in collapsed haloes, and 
the correlation function of the particles depends on the density profiles 
as well as on the spatial distribution of the parent haloes.  
What has changed since those early days is that we now understand that, 
for statistics like the correlation function, the most important parameter 
of a halo is its mass.  

Following Sheth \& Saslaw (1994), Sheth \& Jain (1997) showed that they 
were able to provide a good description of the dark matter correlation 
function on small scales even though they neglected the fact that the 
parent haloes are clustered.  
This works because most close pairs are actually in the same halo, and 
so the correlation function depends only on the distribution of particles 
within haloes (the halo density profile) and not on the spatial 
distribution of the other haloes.  Using formalism presented in 
McClelland \& Silk (1977), they showed how to write the correlation 
function as an integral over haloes having a range of masses.  
They then used simple analytic approximations for $n(m)$, the number 
density of haloes (the mass function formula of Press \& Schechter 1974), 
and halo density profiles (power-laws with slopes chosen to agree with 
simulations), to present their results.  

The results of e.g., Scherrer \& Bertschinger (1991), 
show that one really expects the power-spectrum to be the sum 
$P_{\rm 1halo}(k) + P_{\rm 2halo}(k)$, where $P_{\rm 2halo}(k)$ 
is the contribution from pairs in different haloes (i.e., the term 
that was neglected by Sheth \& Jain).  More recently, Seljak (2000) 
and Peacock \& Smith (2000) have extended the model calculation to 
include these terms.  They used slightly more accurate fitting 
formulae for the inputs (the mass function is from Sheth \& Tormen 1999 
and the halo profiles are from the work of Navarro, Frenk and White 1997).  
Their results show that a good approximation to the evolved $P(k)$ on 
all scales can be got by simply setting 
 $P_{\rm nl}(k) = P_{\rm 1halo}(k) + P_0(k)$, 
where the first term is that due to the halo profiles and mass 
function, and the second is the initial power-spectrum evolved to the 
present time using linear theory.  

Section~\ref{modelxi} of this paper studies the two-point correlation 
function, $\xi(r)$, rather than its Fourier transform, $P(k)$.  
Of course, in this case also, 
$\xi(r) = \xi_{\rm 1halo}(r) + \xi_{\rm 2halo}(r)$, 
but we feel that explicitly working in real space shows the sorts of 
approximations which lead to this decomposition more clearly.  
It also shows why, both in real and in Fourier space, the 2-halo term 
should approximately equal the linear theory expression.  

Since this is a model in which the contributions to $\xi(r)$ are 
written as functions of halo mass, we can study the Layzer-Irvine 
cosmic energy equation as a function of halo mass.  We show this in 
Section~\ref{vrms}.  Our analysis shows which terms in the energy equation 
come from nonlinear virial motions within each halo, and which from the 
motions of haloes as a whole which, following Sheth \& Diaferio (2000), 
are more in line with what the linear theory would predict.  We show how 
our decomposition allows us to provide a simple estimate of the single 
particle velocity dispersion---which can be thought of as the density 
weighted temperature.  

The requirement of pair conservation provides a relation between the 
correlation function $\xi(r)$ and the first moment of the pairwise 
velocity distribution $v_{12}(r)$.  In Section~\ref{v12} we use our 
decomposition of $\xi$ to provide a simple expression for the mean 
streaming velocity, $v_{12}$, and then show that this expression 
describes measurements in simulations quite well. 

Peebles (1980) shows that the second moment, $\sigma_{12}(r)$, is related 
to (an integral over) the three-point correlation function, $\zeta$.  
This relation between the pairwise dispersion and $\zeta$ is often called 
the cosmic virial theorem.  Since our halo-based approach allows us to 
model $\zeta$ as well (Scoccimarro et al. 2000), we could, in principle, 
use this to study how $\sigma_{12}(r)$ depends on pair separation.  
In Section~\ref{s12} we describe what we think is a much simpler 
way to think about and model $\sigma_{12}(r)$.  In our approach, the 
important ingredient is not $\zeta$, but knowledge of how velocities are 
correlated.  We first show that neglecting these correlations is a 
rather good approximation:  our models of $\xi$ and the single particle 
velocity dispersion are sufficient for describing the main features of 
$\sigma_{12}$.  We then describe a simple model for including the 
effects of velocity correlations.  

Knowledge of both the mean and the dispersion of pairwise velocities 
are useful for modelling redshift space distortions, which we plan 
to present elsewhere.  The three velocities we study here, 
the single particle velocity dispersion $\langle v^2\rangle$, 
the mean streaming velocity $v_{12}(r)$, and 
the pairwise dispersion $\sigma_{12}^2(r)$ may all be used to 
provide estimates of the density of the Universe, although 
Jenkins et al. (1998) discuss why, in cluster normalized CDM models, 
different cosmological models may have rather similar values of 
$v_{12}$ and $\sigma_{12}$.

\section{The correlation function}\label{modelxi}
Let $\rho(r|m)$ denote the shape of the density profile of a halo 
which contains mass $m$; the mass in a spherical shell at distance 
$r$ from the centre of the halo is $4\pi r^2\,\rho(r|m)\,{\rm d}r$.  
Let $\lambda(r|m)$ denote the convolution of such a profile with
another of exactly the same shape.  For spherically symmetric density 
profiles 
\begin{equation}
\lambda(r|m)\equiv 2\pi \int {\rm d}x_1\ x_1^2\,\rho(x_1|m)\,
\int_{-1}^1 {\rm d}\beta\ \rho(x_2|m) ,
\label{lambdar}
\end{equation}
where $x_2^2 = x_1^2 + r^2 - 2 x_1 r \beta$.  
The contribution to the correlation function from pairs in which both 
particles are in the same halo is given by weighting the convolution 
profile of a halo of mass $m$ by the number of haloes of mass $m$, and 
integrating over $m$ (Sheth \& Jain 1997).  The total correlation 
function is the sum of this plus a term which arises from pairs which 
are in two different haloes.  This means that the second term is a 
convolution of the profiles of each of the two haloes involved, with 
the halo-halo correlation function:
\begin{eqnarray}
\Lambda(r|m_1,m_2) &=& \int\!{\rm d}^3{\boldmath}r_1 
\int\!{\rm d}^3{\boldmath}r_2 \ \rho(r_1|m_1)\rho(r_2|m_2)\,\nonumber \\
&&\qquad\times\ \ \xi_{\rm hh}\Bigl(|r_1 - r_2 + r|\Big| m_1,m_2\Bigr).
\end{eqnarray}
Suppose that the halo-halo correlation function changes slowly on 
separations which are large compared to the typical size of a halo.  
Then, at large $r$, the halo-halo correlation function can be taken 
outside the integrals, leaving just the convolutions over the profiles.  
But these each contribute a factor which equals the mass of the halo, 
since any position within the halo gives approximately the same pair 
separation.  This means that 
\begin{equation}
\Lambda(r|m_1,m_2) \approx m_1\,m_2\ \xi_{\rm hh}(r|m_1,m_2) 
\end{equation}
at large separations.  

To proceed, we need a model for $\xi_{\rm hh}(r|m_1,m_2)$.  
This has been done by Mo \& White (1996) and Sheth \& Lemson (1999).  
On large scales, Mo \& White argued that the correlation function of 
haloes of mass $m$ should simply be a constant times the correlation 
function of the dark matter, $\xi_{\rm hh}(r) = b^2(m)\,\xi(r)$, and 
that the value of the constant should depend on the halo mass.  
Sheth \& Tormen (1999) showed that this dependence on mass can be 
derived from the shape of the mass function $n(m)$, which, in turn, 
depends on the initial shape of the power spectrum.  
In addition, on large scales, linear theory should apply, 
and so $\xi(r)\approx \xi_0(r)$, where $\xi_0(r)$ is the linear theory 
correlation function of the dark matter.  Thus, for large separations,  
\begin{equation}
\xi_{\rm hh}(r|m_1,m_2) \approx b(m_1)\,b(m_2)\ \xi_0(r).
\end{equation}

On small scales the halo-halo correlation function must eventually turn 
over (haloes are spatially exclusive---so each halo is like a small hard 
sphere).  So setting 
$\xi_{\rm hh}(r|m_1,m_2) \approx b(m_1)\,b(m_2)\,\xi(r)$ 
will almost surely overestimate the true value.  
Using the linear, rather than the nonlinear correlation function, 
even on small scales, is a crude but convenient way of accounting for 
this overestimate.  (Although the results of Sheth \& Lemson 1999 
allow one to account for this more precisely, it turns out that great 
accuracy is not really needed since, on small scales, the correlation 
function is determined almost entirely by the one-halo term anyway.)  

If we allow a range in halo masses then the total correlation function is 
\begin{eqnarray}
\xi(r) &\equiv& \xi_{\rm 1halo}(r) + \xi_{\rm 2halo}(r) \nonumber \\
&=& \int {\rm d}m\,{n(m)\over\bar\rho}{\lambda(r|m)\over\bar\rho} + 
\nonumber \\
&& \int {\rm d}m_1\,{n(m_1)\over\bar\rho} 
\int {\rm d}m_2\,{n(m_2)\over\bar\rho}\ \Lambda(r|m_1,m_2),
\label{xir}
\end{eqnarray}
where $n(m)\,{\rm d}m$ denotes the number density of haloes which have 
mass in the range d$m$ about $m$, and $\bar\rho$ is the average density.  
The first term dominates on small scales, and the second term dominates 
on large separations.  Now, on large scales $\Lambda(r|m_1,m_2)$ 
is well approximated by the product of $m_1\,b(m_1)$, $m_2\,b(m_2)$ 
and $\xi_0(r)$.  We will not do too badly if we continue to use this 
approximation on smaller scales because the scales on which it breaks 
down are precisely those on which the first term begins to dominate.  
This means that the second term can be written as the product of two 
one dimensional integrals.  Moreover, the bias factors are defined so 
that 
\begin{equation}
\int {\rm d}m_1\,{m_1 n(m_1)\over\bar\rho}\, b(m_1) \equiv 1.
\end{equation}
Therefore, the second term really is very simple:  to a good 
approximation, 
\begin{equation}
\xi(r) \approx \int {\rm d}m\,{n(m)\over\bar\rho}\,
{\lambda(r|m)\over\bar\rho} + \xi_0(r).
\label{xitotal}
\end{equation}
Note that, in this approximation, the second term is the same for all 
halo profiles---different profile shapes yield different shapes for 
$\lambda(r|m)$ and so result in different correlation functions on small 
scales only; in principle, one can use measurements of the shape of the 
correlation function at small separations to constrain the shapes of 
profiles.  

To illustrate how the profile affects the correlation function, 
in what follows we will use the NFW profile of Navarro, Frenk \& 
White (1997).  In Appendix~A we provide expressions for 
$\lambda(r|m)$ for the NFW profile truncated at the virial radius, 
the Hernquist profile (1990), and the singular isothermal sphere, 
truncated at the virial radius so that it has finite mass.  These 
three density profiles all have the property that $\lambda(r|m)$ 
is never less than zero.  As a result, $\xi_{\rm 1halo}$ is also 
never less than zero.  This means that the integral of 
$\xi_{\rm 1halo}$ over all separations does not equal zero.  
As a result, equation~(\ref{xitotal}) above does not satisfy the 
integral constraint.  This is a formal feature of the models to 
which we will return later.  

The truncated NFW profile above has two free parameters, a core 
radius and an average density.  Navarro, Frenk \& White (1997) 
showed that, in their simulations, the core radius depends on the 
mass of the halo, whereas all haloes have the same average density, 
whatever their mass.  They found that the core radii of massive 
haloes are at larger fractions of their virial radii than for less 
massive haloes; less massive haloes are more centrally concentrated.
The ratio of the core radius to the virial radius increases with mass 
in a way which depends on the shape of the initial power-spectrum---they 
discuss in detail why this is so.  In what follows, we will use the 
simple analytic approximation to this relation given in Scoccimarro 
et al. (2000) as we integrate $\lambda(r|m)$ over the mass function:
$a_{\rm NFW}(m) = (m/m_*)^{0.13}/9$, where $m_*$ is the average mass 
contained in a tophat filter whose scale is set by the requirement that 
the rms value of the initial density fluctuation field smoothed with the 
filter, extrapolated to the present using linear theory, is 
$\delta_{\rm c}\approx 1.68$.  

\begin{figure}
\centering
\mbox{\psfig{figure=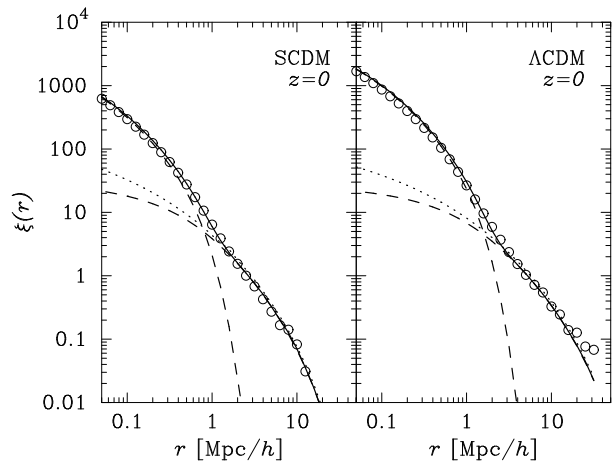,height=6cm,bbllx=72pt,bblly=58pt,bburx=629pt,bbury=459pt}}
\caption{The correlation function of dark matter particles.  
Symbols show the fitting formula provided by Peacock \& Dodds (1996).  
Solid curves show the correlation associated with NFW profiles.  
The two dashed curves show the contribution from pairs in the same 
halo (dominates at small $r$), and pairs in different haloes 
(dominates at large $r$).  Dotted curve shows the linear theory 
correlation function $\xi_0$; on the large scales where the 
two-halo term dominates, it is very well approximated by the linear 
theory function.  }
\label{xiplot}
\end{figure}

Fig.~\ref{xiplot} shows the correlation functions one obtains by 
inserting the core-radius relation given above into our expressions for 
$\lambda(r|m)$, and integrating over the mass function.  The two panels 
show two variants of the CDM family of initial conditions:  our SCDM 
model has $\Omega_0=1$, $h=0.5$ and $\sigma_8=0.5$, whereas our 
$\Lambda$CDM model has $\Omega_0=1$, $\Lambda_0 = 1-\Omega_0$, 
$h=0.7$ and $\sigma_8=0.9$.  
The symbols show the fitting formula of Peacock \& Dodds (1996) which 
describes the correlation functions from numerical simulations of 
these cosmological models well (see, e.g., Mo, Jing \& B\"orner 1997; 
Jenkins et al. 1998), and the solid curve shows our model using NFW 
profiles.  The two dashed curves, which sum to give the solid curve, 
show the two contributions to the correlation function; the curve which 
dominates on small scales shows the contribution from pairs in the same 
halo---the first term in equation~(\ref{xir}).  The curve which dominates 
on large scales shows the contribution from pairs in different haloes.  
The dotted curve shows the linear theory correlation function; on the 
scales where the two-halo term dominates, the linear theory function 
provides an excellent approximation.  The model provides a good description 
of the correlation function of the dark matter.  Although we have 
only presented results for $z=0$, the model is also accurate at earlier 
times.  

\section{The cosmic energy equation}\label{vrms}
The cosmic energy equation describes how the energy of the Universe is 
partitioned between kinetic and potential energy.  
In essence, it provides a relation between the correlation function 
of dark matter particles and their rms speeds.  
Following Hamilton et al. (1991), we know how to compute the shape 
of the correlation function (i.e., how it depends on separation $r$) 
at any time, if its initial shape is known (also see Nityananda \& 
Padmanabhan 1994).  Mo, Jing \& B\"orner (1997) showed that inserting 
this evolution into the cosmic energy equation provided a very good 
description of how the rms velocities of dark matter particles in 
their simulations evolved.  How much of the evolution of the rms 
velocity is driven by nonlinear effects within virialized clusters, 
which produce large velocities, and how much contains information 
about the (linearly evolved) initial conditions?  
We show below that the model described above allows us to separate out 
the linear from the nonlinear effects.  

The cosmic energy equation (Irvine 1961, 1965; Layzer 1963, 1964) is 
an exact statement of what energy conservation means in an expanding 
universe.  It provides a relation between the potential energy of 
the system, 
\begin{equation}
W(a) = -2\pi G\bar\rho(a)\,\int {\rm d}r\ r\,\xi(r,a) ,
\end{equation}
and the kinetic energy $K(a)$, which is essentially one half times 
the mean square velocity of the particles in the system.  (Note that 
$\bar\rho\propto a^{-3}$, where $a$ denotes the scale factor of the 
Universe, and not the core radius of the halo profile!)  If the evolution 
of $W$ is known, then the cosmic energy equation allows one to compute 
the evolution of $K$ also.  Since we know how $\xi(r,a)$ evolves, we 
know how $W$ evolves, so we can compute $K(a)$ also.  
Mo, Jing \& B\"orner (1997) showed that by combining the 
Hamilton et al. (1991) evolution of $\xi(r,a)$ with the cosmic energy 
equation they were able to compute a good approximation to the value of 
the single particle velocity dispersion at any given time.  
In particular, by integrating the energy equation once, 
Mo et al. (1997) showed that 
\begin{equation}
\langle v^2(a)\rangle = {3\over 2}\,\Omega(a)H^2(a)\,a^2 I(a)
\left(1 - \int_0^a {I(a')\over I(a)}\,{{\rm d}a'\over a}\right),
\end{equation}
where $W(a) = -2\pi G\bar\rho(a)\,a^2 I(a)$.  Although this expression 
is exact, Mo et al. also showed that the simpler expression 
which follows if $I(a)\propto D(a)^2$, as it would in linear 
theory, is a good approximation:  In this case
\begin{equation}
\langle v^2(a)\rangle \approx {3\over 2}\,\Omega(a)H^2(a)\,a^2 I(a)
\left(1 - \int_0^a {D^2(a')\,{\rm d}a'\over aD^2(a)}\right).
\end{equation}

Davis, Miller \& White (1997) were interested in separating out those 
contributions to the velocity dispersion which arise from nonlinear 
effects from those which are given by linear theory.  
Since we know how to write $\xi(r,a)$ as a sum of two terms, one 
from linear theory and the other nonlinear, we can begin to address 
some of the issues they raised.  Specifically, suppose we write  
\begin{equation}
W = W_{\rm linear} + W_{\rm nonlinear} 
\end{equation}
and we require that it equals the expression for $W$ above.  
Linear theory, extrapolated to the present time, would have 
\begin{equation}
W(a) = -2\pi G\bar\rho(a)\,\int {\rm d}r\ r\,\xi_0(r),
\end{equation}
where $\xi_0$ denotes the linear correlation function (e.g. Peebles 1980).  
However, our model for the nonlinear correlation function is to write 
$\xi(r)$ as a sum of this linear part, plus a term which depends on 
halo profiles.  Thus, we find that 
\begin{equation}
W_{\rm nonlinear} = -2\pi G\,\int {\rm d}m\ {n(m)\over\bar\rho}\,
\int {\rm d}r\ r\,\lambda(r|m) .
\end{equation}
where we have rearranged the order of the integrals so that we do 
do the one over $r$ before integrating over $m$.  
The resulting integrals over $r$ are computed explicitly in the Appendix.  
In particular, for the halo models of interest in this paper, the haloes 
are in virial equilibrium: $W_{\rm nonlinear} = -2K_{\rm nonlinear}$.  
Thus, in our approach, the ratio of the nonlinear to the linear theory 
term depends on the mass function and the density profiles of dark matter 
haloes.  For example, for NFW haloes, $-W_{\rm nonlinear}$ equals 
$\int\,{\rm d}m\,[n(m)/\bar\rho]\,(Gm^2/r_{\rm vir})$ times a constant 
which depends on the ratio of the core radius to the virial radius.  

We will not explore this further in this paper.  
For the time being, we will simply use this expression 
to estimate how the single particle velocity dispersion evolves.  
Namely, we will use linear theory extrapolated to the present time to 
estimate $W_{\rm linear}$ (see, e.g., Peebles 1980) and our models of 
halo profiles to estimate the other contribution to $W_{\rm nonlinear}$, 
and we will then use the energy equation to derive $\langle v^2\rangle$.  
We will need this single particle $\langle v^2\rangle$ later on, when we 
study how the pairwise velocity dispersion depends on pair separation.  
For the CDM models presented in this paper, this gives 
$\langle v^2\rangle^{1/2}_{\rm SCDM} = 675$km/s and 
$\langle v^2\rangle^{1/2}_{\Lambda\rm CDM} = 590$ km/s, 
in good agreement with the values measured in the simulations 
(e.g. Mo, Jing \& B\"orner 1997; Sheth \& Diaferio 2000).  
Because our models allow us to compute this single particle dispersion 
at any times, they provide a simple way of computing the evolution of 
the density weighted temperature of the Universe.  

\section{The mean streaming velocity}\label{v12}
The scale dependence of the mean streaming $v_{12}(r)$ of 
dark matter particles has been understood for some time now.  
Hamilton et al. (1991) showed that because they could provide good 
estimates of the evolution of $\xi(r,a)$, for any initial correlation 
function, they could also describe the shape of $v_{12}(r)$
(also see Nityananda \& Padmanabhan 1994).  Because the halo model 
presented in the previous section allows one to compute $\xi(r,a)$, 
by following the steps outlined by Hamilton et al., it can also be used 
to compute $v_{12}(r)$.  We will not do this here.  
Rather, we will show that because our halo model allows one to split the 
correlation function up into linear and nonlinear parts, we are able to 
compute a good approximation to $v_{12}(r)$ rather more simply.  
Recently, Juszkiewicz, Springel \& Durrer (1999) have presented a fitting 
formula for this statistic; they argue that their formula is simpler to 
use than the exact method of Hamilton et al.  The results presented in 
this section can be thought of as providing a simple physical reason for 
the values of the coefficients in their fitting formula.  

The relevant starting point is the pair conservation equation in 
Peebles' book (Peebles 1980):
\begin{equation}
 a\, {\partial (1+\bar\xi)\over{\partial a}} = 
-{v_{12}(r)\over Hr}\ 3\,\Bigl[1+\xi(r)\Bigr],
\label{pairs}
\end{equation}
where $\bar\xi(r,a)$ is the volume averaged correlation function 
on comoving scale $x = r/a$ at the time when the expansion factor 
is $a$, and the Hubble constant is $H$.  Note that the partial 
derivative with respect to $a$ on the left hand side keeps $x$ fixed 
rather than $r$.  This says that if we know the 
correlation function for all scales $x$ and all times $a$, then we can 
compute how $v_{12}(r)$ depends on scale today;  basically it comes 
from assuming that the number of pairs is conserved.
Hamilton et al. also showed that by inserting their expression for the 
evolution of $\xi(r,a)$ into equation~(\ref{pairs}) above, they were able 
to describe $v_{12}(r)$ accurately.  

A simpler analytic approach follows from reading through Peebles' 
logic further.  He notes that an approximate solution to 
equation~(\ref{pairs}) can be got by assuming that $\bar\xi$ evolves 
according to linear theory:  
 $\bar\xi(r=ax,a) = [D(a)/D(a_0)]^2 \bar\xi(r=a_0x,a_0)$, 
where $D(a)$ is the linear theory growth factor.  Then the left hand 
side is  
 $a\,\partial \bar\xi(ax,a)/\partial a = 2f(\Omega)\, \bar\xi(ax,a)$, 
where $f(\Omega)\equiv \partial {\rm ln}D/\partial {\rm ln}a 
\approx\Omega^{0.6}$.  So, in this approximation we get 
\begin{equation}
-{v_{12}\over Hr} = 
{2\over 3}{f(\Omega)\,\bar\xi(r,a)\over 1 + \xi(r,a)} 
\label{peeb}
\end{equation}
This is just the usual linear theory expression with an extra factor 
of $(1 + \xi)$ in the denominator.  Juszkiewicz et al. (1999) show that, 
while this approximation is fine on large scales, it underestimates the 
exact solution by a factor of 3/2 or so on smaller scales.  They use 
perturbation theory to motivate the introduction of the extra terms 
they must add to this expression to rectify this problem.  

Our model provides another simple way to see what these terms should be.  
Since the previous section allows us to write $\xi$ as a sum of two terms, 
we can work out how each one scales with time.  In our model, the term 
which dominates on small scales evolves both because the mass function 
evolves, and because the concentrations of haloes of a fixed mass depend 
on when they formed.  The term which dominates on larger scales 
is very similar to that predicted by linear theory.  If we assume that 
it evolves according to linear theory, then 
\begin{equation}
-{v_{\rm 12}\over Hr} = {1\over 3[1+\xi(r,a)]} 
\left[ 2 f(\Omega) \bar\xi_{\rm 2halo} (r,a) + 
{\partial \bar\xi_{\rm 1halo} \over \partial {\rm ln} a}\right],
\label{v12model}
\end{equation}
where the derivative of $\xi_{\rm 1halo}$ with respect to 
expansion factor $a$ can be evaluated because the dependence of the 
mass function $n(m,a)$ and the halo profiles and their convolutions 
$\lambda(r|m)$ on $a$ are all known.  Thus, we have 
\begin{eqnarray}
{\partial \bar\xi_{\rm 1halo} \over \partial {\rm ln} a} &=& 
{\partial {\rm ln} m_* \over \partial {\rm ln} a}
\Bigl[\bar\xi_{\rm 1halo} (r,a) - \xi_{\rm 1halo} (r,a)\Bigr] \nonumber \\
&&\ \ + \ {3\over r^3} \int_0^r {\rm d}r'\,{r'}^2
\int_0^\infty {\rm d}m {n(m)\over\bar\rho} \nonumber\\
&&\qquad \times \ {\lambda(r|m)\over\bar\rho}
{\partial {\,\rm ln}\lambda \over \partial {\,\rm ln}c}
\left. {\partial {\,\rm ln}c \over\partial {\,\rm ln}a}\right |_{m \over m_*}.
\label{v12profile}
\end{eqnarray}
Here $c$ denotes the inverse of the core radius relation of the halo 
profile (so $c_{\rm NFW}=1/a_{\rm NFW}$) to avoid confusing the core-radius 
relation with the cosmological expansion factor.  The notation 
$\left. {\partial {\,\rm ln}c / \partial {\,\rm ln}a} \right |_{m/m_*}$ 
denotes a derivative with respect to ${\,\rm ln}a$ keeping $m/m_*$ fixed.
If the time dependence of $c$ comes only from its dependence on 
$m_*$, then the final term on the right hand side vanishes,
making the expression particularly simple.  If, in addition, 
the correlation function was a pure power-law of slope, say, $\gamma$, 
then $\xi = (3-\gamma)\bar\xi/3$.  In this case, the term in square 
brackets in the expression above would become 
$(\gamma/3)\,\bar\xi_{\rm 1halo}$.  

To illustrate how the one-halo term scales, it is convenient to study 
a spectrum with shape $P(k)\propto k^n$ initially.  In this case 
$\partial {\rm ln} m_*/\partial {\rm ln} a = f(\Omega)\,6/(3+n)$.  
On very small scales, $\xi\approx\xi_{\rm 1halo}$ is expected to be 
a power law of slope $\gamma_{\rm SC}\approx 3(3+n)/(5+n)$ 
(e.g. Peebles 1980).  This makes the right hand side of 
equation~(\ref{v12profile}) equal to $f(\Omega)\,6/(5+n)$ times 
$\bar\xi_{\rm 1halo}$.  This is the same as the scaling required by 
stable clustering (e.g. Hamilton et al. 1991).  
On intermediate scales, Padmanabhan (1996) argues that $\bar\xi$ 
should be approximately a power-law of slope 
$\gamma_{\rm QL}\approx 3(3+n)/(4+n)$.  If we still set 
$1+\bar\xi\approx \bar\xi_{\rm 1halo}$, then the right hand side 
of equation~(\ref{v12profile}) becomes $f(\Omega)\,6/(4+n)$ times 
$\bar\xi_{\rm 1halo}$.  This is precisely the quasi-linear scaling 
assumed by Padmanabhan (1996).  
On large scales $\bar\xi$ has slope $\gamma_{\rm L} = (3+n)$, and so 
the right hand side of equation~(\ref{v12profile}) becomes 
$2\,f(\Omega)\,\bar\xi_{\rm 1halo}$, which is the same as the scaling 
required by linear theory.  Thus, our halo profile term 
(equation~\ref{v12profile}) interpolates smoothly between these 
different regimes.  Of course, because $\xi_{\rm 1halo}(r)$ is not 
really a power law, it never actually obeys these scalings exactly.  
Hamilton et al. (1991) assumed that $v_{12}$ could be written as a 
function of $\bar\xi$ alone.  Because our one-halo term actually depends 
both on $\bar\xi_{\rm 1halo}$ as well as on $\xi_{\rm 1halo}$, our halo 
models are formally inconsistent with the Hamilton et al. ansatz.  
For example, the ansatz is based on an assumption that clustering on 
small scales is stable, whereas our halo models are not.  
Ma \& Fry (2000) explore some consequences of this.  

\begin{figure}
\centering
\mbox{\psfig{figure=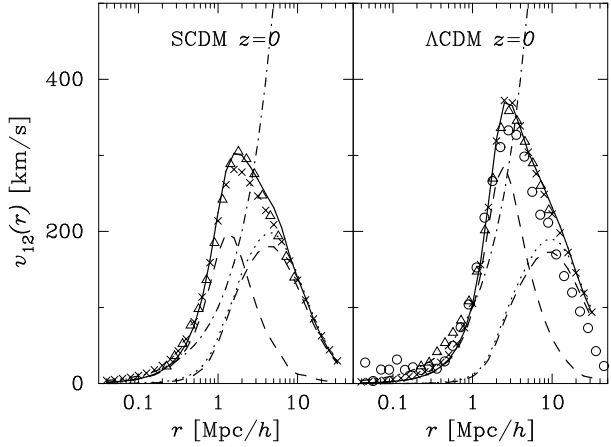,height=6cm,bbllx=72pt,bblly=58pt,bburx=629pt,bbury=459pt}}
\caption{The mean streaming velocity of dark matter particles.   
Triangles show the Virgo simulation measurements, 
circles show the GIF $\Lambda$CDM simulation, 
and dot-dashed curves show the Hubble expansion velocity.  
Crosses show the result of using the Peacock \& Dodds (1996) formulae 
for the correlation function in the fitting formula provided by 
Juszkiewicz et al. (1999).  
Solid curves show the model described in the text which accounts for 
the fact that the nonlinear evolution is different from what linear 
theory predicts, and then weights the linear and nonlinear scalings 
by the relative fractions of linear and nonlinear pairs.
Dashed curves show the two contributions to the streaming motion 
in our model; the curves which peak at large $r$ are for pairs in 
two different haloes.  Dotted curve shows the approximation of using 
the linear theory correlation function to model this two-halo term. }
\label{v12plot}
\end{figure}

Fig.~\ref{v12plot} shows that our model, 
\begin{eqnarray}
-{v_{\rm 12}\over Hr} &=& {f(\Omega)\over 3[1+\xi(r,a)]} 
\Biggl[ 2\bar\xi_{\rm 2halo} (r,a) + \nonumber \\
&&  {6\over 3 + n_*}
\Bigl[\bar\xi_{\rm 1halo} (r,a) - \xi_{\rm 1halo} (r,a)\Bigr]\Biggr].
\end{eqnarray}
where $n_*=-1.33, -1.53$ is the slope of the power spectrum on the 
scale $m_*$ for the SCDM and $\Lambda$CDM models we consider in this 
paper, is quite accurate.  The triangles show measurements from the 
publically available Virgo simulations (Jenkins et al. 1998), and 
crosses show the fitting formula which Juszkiewicz et al. (1999) 
obtained by fitting to these simulations.  The open circles in the 
panel on the right show the streaming motions in the $\Lambda$CDM GIF 
simulation (e.g. Kauffmann et al. 1999).  The GIF SCDM box is only 
85 Mpc$/h$ on a side.  As a result, the velocities in it are strongly 
affected by the finite size of the box (Sheth \& Diaferio 2000), which 
is why we have not shown $v_{12}$ from this simulation.  The solid 
curve shows our model (equation~\ref{v12model}) which accounts for the 
fact that evolution on small scales is nonlinear, and weights by the 
relative fractions of linear and nonlinear pairs.  It is the sum 
of two terms; these are shown as the two dashed curves.  Using the 
linear theory correlation function (the dotted curve in the previous 
figure) to model the contribution from the two-halo term (the dashed 
curve which dominated on large scales in the previous figure) corresponds 
to setting $\bar\xi_{\rm 2halo}\to \bar\xi_0$.  This approximation is 
shown as the dotted curves.  The dot-dashed curves show the Hubble 
expansion velocity for comparison.  

\begin{figure}
\centering
\mbox{\psfig{figure=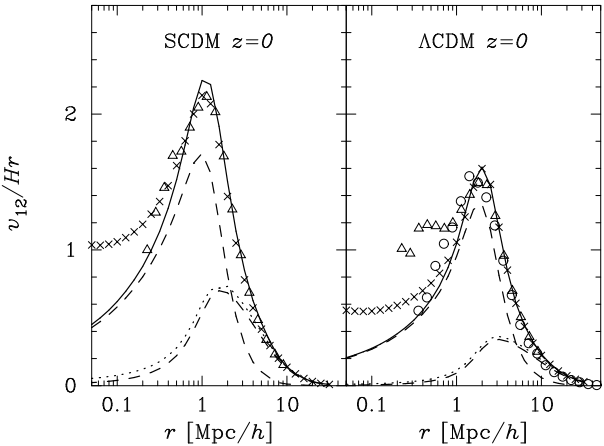,height=6cm,bbllx=72pt,bblly=58pt,bburx=629pt,bbury=459pt}}
\caption{The ratio of the mean streaming velocity of dark matter 
particle pairs separated by $r$, to the Hubble expansion on that 
scale.  As in the previous Figure, triangles show the Virgo 
simulations, circles show the $\Lambda$CDM GIF simulation, 
crosses show the Juszkiewicz et al. (1999) fitting formula, and solid 
curves, which are the sum of the dashed curves, show our model 
predictions.}
\label{v12Hr}
\end{figure}

To study the approach to stable clustering, it is more useful to 
show the ratio of the streaming velocity to the Hubble expansion.  
Recall that one might have expected the smallest scales to have 
$-v_{12}/Hr = 1$.  This follows from the simple scalings 
we discussed earlier (assuming $\Omega_0=1$): 
set $\bar\xi_{\rm 1halo} = 3\xi_{\rm 1halo}/(3-\gamma_{\rm SC})$, 
and use the fact that $1+\xi\approx\xi_{\rm 1halo}$.  
Fig.~\ref{v12Hr} shows that on small scales the mean streaming motions 
in our halo models are smaller than the stable clustering limit.  The 
streaming motions do not cancel the Hubble expansion.  As before, 
the triangles in the two panels show the Virgo simulations, the
crosses show the fitting formula to these simulations, and circles 
in the panel on the right show the streaming motions in the 
$\Lambda$CDM GIF simulation.  Although our model $v_{12}$ curves 
(solid lines) fall below the stable clustering limit, they are in 
quite good agreement with the simulations.  

Notice that, on scales slightly smaller than a Mpc$/h$ or so, the 
one-halo term in our models can exceed the Hubble velocity.  
This shows explicitly that, even if clustering were to approach the 
stable clustering limit, it would have to do so on scales which are 
smaller than the virial radii of haloes.  This is consistent with 
arguments in Sheth \& Jain (1997).  Another way of saying this is that,
if $-v_{12}$ equalled $Hr$ out to the virial radii of all haloes, then 
the one-halo contribution to the mean streaming velocity would be 
$\xi_{\rm 1halo}/(1 + \xi)$:  this would give a curve which decreased 
monotonically from unity as $r$ increased, whereas our one-halo term 
actually has a peak at $\sim 1$Mpc$/h$.  On these slightly larger 
scales, writing our previous scalings for power-law profiles in terms 
of $\xi$ with slope $\gamma_{\rm QL}$ shows that 
$-v_{12}/Hr = 2f(\Omega)$.  
This value is in reasonable agreement with the height of the peak of 
the solid curve in both panels of Fig.~\ref{v12Hr}.  
Higher resolution simulations are required to determine whether 
clustering is stable on scales smaller than about $0.1$Mpc$/h$.

The careful reader will have noticed that our halo models slightly 
overestimate the amplitude of the streaming motions on large scales.  
Some of this effect may be due to our simple treatment of 
$\xi_{\rm 2halo}$.  If we were to set $\xi_{\rm 2halo}\to\xi_0$ 
we would overestimate the true value even more (the dotted curves 
are always above the dashed ones).  However, notice that the 
single-halo contribution to $v_{12}$ is nonzero even at separations 
larger than 10Mpc$/h$, and that it approximately accounts for the 
overestimate on these scales.  Where does this large scale single-halo 
contribution come from?  It does not arise from particles which are 
falling towards each other from opposite sides of a few proud monster 
haloes!  (A virialized halo with a radius of 5 Mpc$/h$ would have a mass 
of about $2.5\Omega\times 10^{16}M_\odot/h$.)  Rather, at least some of 
the overestimate of $v_{12}$ arises from the fact that, formally, the 
halo models violate the integral constraint.  As mentioned earlier, the 
integral of the halo model correlation function over all separations does 
not equal zero.  This means that the model overpredicts the value of 
the true $\bar\xi_{\rm 1halo}$.  A glance at Fig.~\ref{xiplot} shows 
that $\xi_{\rm 1halo}$ on the scale of 5 Mpc$/h$ is negligible.  This 
in equation~(\ref{v12profile}) shows that, on these scales, the 
single-halo contribution to $v_{12}$ is determined almost entirely by 
the volume average term, and this results in an overestimate.  

One way of remedying this is to make the halo profiles compensated, say, 
by embedding them in slight underdensities.  This would have the effect 
of making the large scale value of $\xi_{\rm 1halo}$ go slightly 
negative, with a more dramatic effect on large scale values of the 
volume average $\bar\xi_{\rm 1halo}$.  While compensated profiles may 
be physically reasonable, and are certainly of formal interest, we have 
not pursued this further, because the large scales where the spurious 
streaming velocities appear are also those where the two-halo term 
dominates the pair statistics.  

Before we move on to the second moment of the pairwise velocity 
distribution, we think it is useful to rewrite our expression for 
the mean streaming motions one final time.  If we use the fact 
that $\bar\xi_{\rm 2halo}\approx \bar\xi_0$, then 
\begin{eqnarray}
-{v_{\rm 12}\over Hr} &\approx& -{v^{\rm Lin}_{\rm 12}\over Hr}\,
{1+\xi_0(r)\over 1+\xi(r)}  + \nonumber \\
&& \qquad 
{2f(\Omega)\over 3 + n_*}
\Bigl[{\bar\xi_{\rm 1halo}(r)\over \xi_{\rm 1halo}(r)} - 1\Bigr]\,
{\xi_{\rm 1halo} (r)\over 1+\xi(r)},
\label{v12decomp}
\end{eqnarray}
where we have omitted writing factors of $a$ throughout for 
brevity, and we have defined 
$-v^{\rm Lin}_{\rm 12}/Hr \equiv 2\bar\xi_0/3[1 + \xi_0]$, which 
can be written entirely in terms of linear theory quantities.  
The factors $\xi_{\rm 1halo}/[1+\xi]$ and $[1+\xi_0]/[1+\xi]$ 
are simply the fractions of pairs from particles in the same halo, 
and in separate haloes, respectively.  This form shows clearly that 
the streaming motions arise from applying linear theory to the 
pairs in separate haloes, nonlinear theory to the pairs in the 
same halo, and weighting by the fraction of pairs of each type.  
For instance, stable clustering would yield unity times the 
nonlinear pair-weight term, and the quasi-linear scaling described 
above would yield two times the nonlinear pair-weight term.  
It is this sort of decomposition into linear and nonlinear parts 
which we will exploit in what follows.  

\section{The pairwise velocity dispersion}\label{s12}
This section provides a simple model of how and why the pairwise 
velocity distribution depends on scale.  The model constructed 
here is a natural generalization of that studied by Sheth (1996), 
Diaferio \& Geller (1996), and Sheth \& Diaferio (2000).  
It is much simpler than the exact `cosmic virial theorem' approach 
(Peebles 1980; Bartlett \& Blanchard 1995; Mo, Jing \& B\"orner 1997)
one is led to if one attempts to climb the rungs of the BBGKY 
hierarchy, one-by-one.  

\subsection{The model assumptions}
Let $u(r_{12})$ denote the difference between the velocities of two 
particles separated by $r_{12}$, along their line of separation.  
If the velocities of the particles separated by $r_{12}$ are independent 
of each other, then the shape of this distribution can be computed from 
knowledge of the shape of the single particle distribution function 
directly.  On small scales, this assumption of independence is almost 
certainly wrong.  However, in the small separation limit, progress can be 
made by assuming that all pairs at a given small $r_{12}$ are in the same 
halo.  Since haloes are virialized, particle velocities within a halo 
are drawn from independent Maxwellian distributions, and so velocity 
differences are also Maxwellian, albeit with twice the dispersion of the 
single particle case.  The dispersion of the Maxwellian depends on the 
mass of the parent halo in which the pair is, and so the full 
distribution of $u(r_{12})$ can be computed by integrating over the 
distribution of halo masses, weighting by the number of pairs which 
have separations $r_{12}$ within each halo of mass $m$.  This model 
is studied in detail in Sheth (1996) and Diaferio \& Geller (1996).  
Of course, their model only applies on scales where, for most pairs, 
both members are in the same halo.  

What happens at larger separations?  In this subsection we will 
study the limit in which both members of the pair are in different 
haloes.  We will argue that, in this limit, the distribution of $u$ 
is also relatively simple.  

As for the correlation function, we begin by writing the dispersion 
$\sigma_{12}^2(r) = \langle u^2(r)\rangle$ at separation $r$ as the sum 
of two terms:  
\begin{equation}
\sigma_{12}^2(r) = \sigma_{\rm 1halo}^2(r) + \sigma^2_{\rm 2halo}(r),
\label{sig12r}
\end{equation}
where the first term arises from pairs in which both members are in 
the same halo (so it depends on properties of virialized haloes and 
dominates on small scales), and the second term is for pairs in different 
haloes (so it dominates on large scales).  

As mentioned above, the first term depends on the density profiles of 
virialized haloes.  There are two reasons why it is a function of scale.  
First, the velocity dispersion within a halo may depend on position 
within it, so the pairwise dispersion will depend on separation.  However, 
the pairwise dispersion will depend on scale even if we neglect this 
dependence.  To see why, note that we should get a reasonable estimate 
of this term by setting the dispersion equal to the circular velocity at 
the outside edge of the halo:  $Gm/r_{\rm vir}$ (this would be exact for 
an isothermal sphere, but is only approximate otherwise), and using this 
for all $r\le r_{\rm vir}$.  In this case, 
\begin{equation}
\sigma_{\rm 1halo}^2(r) = {1\over 1+\xi(r)}\int {\rm d}m \,
{Gm\over r_{\rm vir}}\,{n(m)\over\bar\rho}\,{\lambda(r|m)\over\bar\rho}.
\label{sig12p}
\end{equation}
This expression is exactly like the one in Sheth (1996), except that 
we have multiplied it by an additional factor of $\lambda(r|m)/[1+\xi(r)]$ 
to account for the fact that only a fraction of the pairs at a given 
separation $r$ are in the same $m$-halo.  

To see that this is the correct pair weighting factor, let 
$U(1,2)\,{\rm d}v_1{\rm d}v_2$ denote the probability that there exists 
a particle in a cell of size $dv_1$ within a halo of mass $m$, and another 
particle in a cell of size $dv_2$ which is not necessarily in the same 
halo.  Similarly, let $P(1,2|m)\,{\rm d}v_1{\rm d}v_2$ denote the 
probability that there exists a particle in a cell of size $dv_1$ within 
a halo of mass $m$, and that the other particle in $dv_2$ is within the 
same halo.  If each particle carries a mass $m_p$, then 
$U = (\bar\rho/m_p) {\rm d}v_1 (\bar\rho/m_p)\,[1+\xi(r)]\,{\rm d}v_2$, 
and $P = mn(m){\rm d}m\,{\rm d}v_1/m_p\, [\lambda(r|m)/m]\,{\rm d}v_2/m_p$.
The ratio is the fraction of pairs at separation $r$ which are both in 
the same $m$-halo:  
$P/U = {\rm d}m\,n(m)\lambda(r|m)/(\bar\rho/m_p)^2 [1+\xi(r)]$.
This is the weighting we used in the equation above.  
Note that this is consistent with equation~(\ref{xitotal}):  one plus 
the left hand side of that expression is the total number of pairs at 
separation $r$, and the first term on the right hand side is the 
contribution from pairs in which both particles were in the same halo.  
It is also the weighting associated with our final expression for 
the mean streaming motions (equation~\ref{v12decomp}).

Allowing the dispersion to depend on position in the halo means that 
\begin{eqnarray}
{Gm\over r_{\rm vir}}\,\lambda(r|m) &\to& 
2\pi\int {\rm d}x_1\ x_1^2\,\rho(x_1|m)
\int_{-1}^1 {\rm d}\beta\,\rho(x_2|m)\nonumber \\
&&\qquad \times \ \ 
\Bigl[\sigma_{\rm 1d}^2(x_1|m) + \sigma_{\rm 1d}^2(x_2|m)\Bigr],
\end{eqnarray}
where $x_2^2 = x_1^2 + r^2 - 2 x_1 r \beta$, and where the one 
dimensional dispersion, $\sigma_{\rm 1d}^2$, can be approximated by, 
say, the radial velocity dispersion given in Appendix~A.

This shows that at small separations, $\sigma^2_{\rm 1halo}(r)$ 
increases with $r$ because virial motions within haloes increase as 
$m^{2/3}$, and only massive haloes can contribute pairs at moderately 
large separations (there will be an additional scale dependence if we also 
included the dependence of the dispersion on position within the halo).  
Therefore the pairwise dispersion will increase with increasing scale 
at small $r$.  On larger scales, however, an increasing fraction of 
pairs are actually from different haloes.  Since $\lambda(r|m)\to 0$ as 
$r$ increases, on scales larger than that of a typical halo, 
$\sigma^2_{\rm 1halo}(r)$ will eventually decrease.  

The term in which particles are in different haloes is 
\begin{eqnarray}
\sigma^2_{\rm 2halo}(r) &=& \int\!\!\int {\rm d}m_1 {\rm d}m_2 
{1 + \xi(m_1,m_2|r)  \over [1 + \xi(r)]} \nonumber\\
&&\ \ \times \ \ 
{m_1 n(m_1)\over\bar\rho}\,{m_2 n(m_2)\over\bar\rho}\,S(m_1,m_2|r),
\label{sig12l}
\end{eqnarray}
where $n(m)$ is the number density of $m$-haloes,
\begin{equation}
\xi(m_1,m_2|r) = b(m_1)b(m_1)\,\xi_0(r)
\end{equation}
is the correlation function of haloes, $b(m)$ is the linear bias factor 
discussed earlier, $\xi(r)$ is the correlation function of the mass, 
and $S(m_1,m_2|r)$ represents the dispersion of the velocity difference
along the line of separation between particles separated by $r$ that are 
in different haloes (one of mass $m_1$ and the other $m_2$).  Whereas the 
other terms are simply the pair-weighting, the physics is in finding a 
convenient expression for $S$.  

If $\sigma^2(m)$ denotes the dispersion associated with a single halo, 
then 
\begin{equation}
S(m_1,m_2|r) = \sigma^2(m_1) + \sigma^2(m_2) - 2\Psi(m_1,m_2|r), 
\label{shalor}
\end{equation}
where $\Psi(m_1,m_2|r)$ represents the fact that the motion of a 
particle in halo $m_1$ may be correlated with that of the particle 
a distance $r$ away in the halo $m_2$.  Later in this paper we will develop 
a model for the velocity correlation function.  For the time being, we think 
it clearer to study a simpler case first---in what follows, we will neglect 
the fact that halo velocities may be correlated.  

Before we do so, notice that if haloes were spatially uncorrelated as well, 
then the pairwise distribution would be the difference of two random 
variates, so it would be a simple convolution of the single particle 
distribution derived by Sheth \& Diaferio (2000).  In what follows, we will 
study what happens if we account for the fact that haloes are spatially 
correlated, even though we neglect the fact that their velocities are 
also correlated.  

\subsection{Neglecting velocity correlations}
Suppose we wish to compute the one-dimensional relative pairwise velocity 
dispersion along the line of separation.  Since we are ignoring velocity 
correlations, the shape of the pairwise velocity distribution arises 
from applying the appropriate pair-weight as a function of separation, 
and integrating over the distribution of dispersions given by the halo 
mass function (equation~(\ref{sig12l}) with $\Psi=0$).  

The pair-weighting involves the bias factor which depends on 
halo mass and on the shape of $n(m)$; since we are using the mass function 
given in Sheth \& Tormen (1999), we will use their formula for $b(m)$ also. 
Let $\langle v^2\rangle$ denote the single particle velocity dispersion 
(section~\ref{vrms} showed that it has contributions from the virial 
motions within haloes as well as from the motions of the haloes 
themselves).  Then the bias weighting means that we must compute 
$2\,\langle b\,v^2\rangle$.  Thus, we arrive at a simple expression for 
the scale dependence of this second term:
\begin{equation} 
\sigma^2_{\rm 2halo}(r) = {2\over 3}\,\langle v^2\rangle\ 
{1 + \xi_0(r) \langle b\,v^2\rangle/\langle v^2\rangle\over 1 + \xi(r)},
\end{equation}
where $\langle v^2\rangle$ is the three dimensional dispersion given by 
the cosmic energy equation.  The factor of $2/3$ arises because we are 
interested in the sum of two (assumed independent) velocity variates, 
and we are only interested in one of the three velocity components.  

At large separations $\xi(r)\approx \xi_0(r)$, and they are both $\ll 1$, 
so $\sigma^2_{\rm 2halo}\to 2\, \langle v^2\rangle/3$, as one expects 
for independent variates.  What about intermediate separations?  
In this model, $\sigma^2_{\rm 2halo}(r)$ depends on scale primarily 
because $\xi(r)$ does.  On intermediate scales (where the assumption 
that the velocities are independent is certainly wrong!), whether or not 
$\sigma^2_{\rm 2halo}$ increases or decreases with scale depends on the 
ratio $\langle b\,v^2\rangle/\langle v^2\rangle$.  For example, the 
weighting by $b$ increases the contribution from massive haloes relative 
to less massive ones.  Since virial motions within massive haloes generate 
velocities that are larger than the rms, for the dark matter, this ratio 
is likely to be larger than unity.  This means that $\sigma^2_{\rm 2halo}$ 
may be slightly higher on intermediate scales than on large ones.  
On small scales, the pair-weighting term $\xi_0/(1+\xi)$ decreases 
rapidly, so $\sigma^2_{\rm 2halo}$ will also decrease.  

\begin{figure}
\centering
\mbox{\psfig{figure=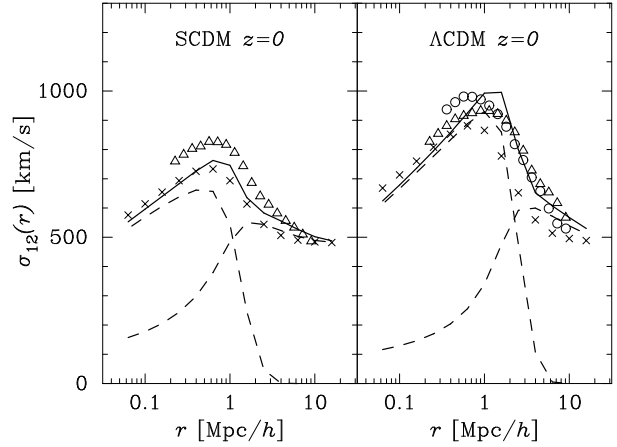,height=6cm,bbllx=72pt,bblly=58pt,bburx=629pt,bbury=459pt}}
\caption{Scale dependence of the pairwise velocity dispersion.  
Triangles show the Virgo simulation results of Jenkins et al. (1998), 
circles show the GIF simulation results, and crosses show the fitting 
formula for the dark matter provided by Mo, Jing \& B\"orner (1997).
Solid curves show the scale dependence in our model which neglects the 
spatial dependence of the dispersion within a halo, and also neglects 
velocity correlations between haloes, but includes the effects of spatial 
correlations.  Dashed curves show the contribution from pairs in which 
both particles are in the same halo (hence the peak at small $r$) and in 
separate haloes (so a peak at large $r$), respectively.  }
\label{sig12}
\end{figure}

Fig.~\ref{sig12} compares our simple model with what is measured 
in simulations.  
Circles show the GIF simulation results, and triangles show 
the VIRGO simulation results of Jenkins et al. (1998).  Crosses show 
a fitting function, equation~(40) of Mo, Jing \& B\"orner (1997), 
which should describe the scale dependence of the pairwise dispersion.  
In all cases, we have chosen to follow standard practice and not
centre the statistic:  to centre, the mean streaming motion $v_{12}$ 
should be subtracted in quadrature.  

Although the GIF and Virgo simulation results are in reasonable 
agreement on large scales, $\sigma_{12}^{\rm GIF}$ is larger, by 
about 100 km/s, on scales smaller than a Megaparsec or so.  Presumably, 
this difference arises from the fact that, although the two simulations 
used the same number of particles, the boxes had different sizes and 
the particle masses were also different.  
The Virgo boxes had sides of length $L=240$Mpc$/h$, 
the GIF $\Lambda$CDM box was 141 Mpc$/h$.  The finite box size has 
two effects:  first, large scale flows have smaller amplitudes in small 
boxes, and so this affects the large-scale value of the pairwise 
dispersion.  In addition, small boxes do not have a fair sample of the 
massive haloes which dominate the pairwise velocity statistic at a 
Megaparsec.  Because massive haloes are rare, the pairwise statistic 
will have a large scatter; the large value of $\sigma_{12}$ we measured 
in the smaller GIF simulation may simply be a large fluctuation.  
On the other hand, because of its better mass resolution, the GIF box 
is able to resolve substructure within virialized haloes which the Virgo 
simulations can not.  The presence of substructure will increase the 
velocity dispersion.  However, because this contribution only adds in 
quadrature, it is not clear that this can account for all the difference. 

We have tried to incorporate the effect of the finite box size into our 
model by restricting our integrals over halo masses to the range 
$m<10^{16} M_\odot/h$, and by only integrating over $k>2\pi/L$ when 
using the cosmic energy equation to estimate $\langle v^2\rangle$.  
The dashed curves show the two types of terms in our model after 
setting the box size to $L=141$Mpc$/h$.  
We chose to model the smaller box because the GIF semianalytic galaxy 
formation models of Kauffmann et al. (1999), which we will use later in 
this paper, use this same simulation.  
The dashed curve which peaks at small $r$ is for pairs in which both 
particles are in the same halo, and the dashed curve which peaks at 
large $r$ is for particles in different haloes.  The solid curve is the 
sum (in quadrature) of the two contributions, plus a piece which 
comes from the fact that the statistic is not centred (see below).  
Our model appears to describe the main features of $\sigma_{12}(r)$ 
reasonably well.  

On small scales, discrepancies between the model and simulations 
are not caused by our neglect of the halo velocity correlation function; 
this regime is dominated by pairs which are in the same halo, which 
suggests that it is the assumption that the pairwise dispersion within 
haloes is isotropic and independent of position within the halo, 
as it would for an isothermal sphere, which should be changed.  
If one is willing to assume the orbits within the halo are isotropic, 
then this can be done by using the expressions for the radial velocity 
dispersion we provide in Appendix~A.  We will show the results of doing 
this in the next subsection.  

On large scales, the difference between the symbols and the solid curve 
is a measure of the importance of velocity correlations.  The relatively 
good agreement on larger scales suggests that our neglect of the halo 
velocity correlation function is not a bad approximation.  
We think it remarkable that we are able to provide a reasonably accurate 
description of the rise and fall of the pairwise velocity dispersion 
without once mentioning the three-point correlation function.  

\subsection{Including velocity correlations}
This subsection shows how one might include the effects of correlated 
velocities.  To do so, we briefly summarize the results of Sheth \& 
Diaferio (2000).  They argued that in a model in which all particles 
are in haloes, such as the one we are studying in this paper, it 
is sensible to write a particle's velocity as the sum of two terms:  
\begin{equation}
v = v_{\rm vir} + v_{\rm halo},
\end{equation}
where $v_{\rm vir}$ is the virial motion of the particle about the halo 
centre of mass, and $v_{\rm halo}$ is the motion of the parent halo.  
Let $\sigma^2(m)$ denote the dispersion of particle velocities in 
$m$-haloes.  It can be written as the sum of the two terms: 
$\sigma_{\rm vir}^2(m) + \sigma_{\rm halo}^2(m)$.  
Sheth \& Diaferio showed that $\sigma^2_{\rm vir}\propto Gm/r_{\rm vir}$,
and that (appropriately smoothed) linear peak theory (Bardeen et al. 1986) 
could be used to estimate $\sigma_{\rm halo}^2(m)$ at any given time 
rather accurately.  Namely, 
$\sigma_{\rm halo}(m) \approx H\Omega^{0.6}\sigma_{-1}(m)\,C(m)$, where 
$\sigma_{-1}$ is computed by multiplying $P(k)$ with a smoothing filter 
$W[kR(m)]$ of scale $R(m)$, integrating over $k$, and dividing by 
$2\pi^2$, and $C(m)$ is a correction factor which accounts for the fact 
that peaks have slightly lower rms velocities than random patches 
(see Sheth \& Diaferio for the exact expressions).   

In such a model, virial motions are random:  the virial velocity of a 
particle is not correlated with the motion of the other particles in the 
halo, nor with the value of $v_{\rm halo}$, nor with the motion of any 
other particle in any other halo.  This means that correlated motions 
arise because halo motions may be correlated, and not otherwise.  
If the virial velocity is independent of pair separation $r$, then 
$\Psi(m_1,m_2|r)$ in equation~(\ref{shalor}) represents the correlations 
of halo motions for $m_1$ and $m_2$-haloes separated by $r$.  Strictly 
speaking, a range of halo separations can contribute to the same particle 
separation $r$, but we are ignoring that here, just as we did when 
computing the correlation function.  

Since linear theory predicts how the velocity correlation function depends 
on smoothing scale (G\'orski 1988),  and since linear theory provides 
a reasonably good description of the rms motions of haloes (Sheth \& 
Diaferio 2000), it is relatively straightforward to include 
the linear theory correlations in our model.  
The final term comes from the fact that the statistic is not centred.  

We will approximate the velocity correlation as follows.  
Suppose for the time being that ignoring the peak constraint gave a 
reasonable approximation to halo velocities.  Then  
$\sigma_{\rm halo}(m) \approx H\Omega^{0.6}\sigma_{-1}(m)$.  
Similarly, the correlation in velocity between patches of different sizes, 
say $R(m_1)$ and $R(m_2)$, along the line of their separation, 
is 
\begin{equation}
\psi(m_1,m_2|r) \equiv H^2\Omega^{1.2}
\int\!\!{{\rm d}k\over 2\pi^2}P(k)\,W(k|m_1,m_2)\,K(kr),
\end{equation}
where 
\begin{displaymath}
W(k|m_1,m_2) \equiv W[kR(m_1)]\,W[kR(m_2)],
\end{displaymath}
and the $W$s are the Fourier transforms of tophat window functions, 
and
\begin{displaymath}
K(x) = {\sin\,x\over x} - {2\over x^3}(\sin\,x - x \cos\,x).
\end{displaymath}
This generalizes the expression in G\'orski (1988), which assumed 
$R(m_1)=R(m_2)$.  
A simple way to include the peak constraint, at least approximately, 
is to multiply by the appropriate peak constraint factors:
\begin{equation}
\Psi(m_1,m_2|r) \approx {\sigma_{\rm halo}(m_1)\over\sigma_{-1}(m_1)}\,
{\sigma_{\rm halo}(m_2)\over\sigma_{-1}(m_2)}\ \psi(m_1,m_2|r).  
\label{psill}
\end{equation}
This includes the fact that peaks have lower rms velocities than 
random patches, but assumes that the correlations are otherwise 
unchanged. 

At first sight, inserting this correlation term into 
equation~(\ref{sig12l}) appears to involve a triple integral---one over 
$m_1$, one over $m_2$, and one over $k$.  In practice, it is much more 
efficient to rearrange the order of the integrals so that the integral 
over $k$ is done last.  The integrals over $m_1$ and $m_2$ are separable 
and equal.  If we use $\Sigma(k)$ to denote the result of the integral 
over $m$ of $mn(m)$ times the window function $W[kR(m)]$, and 
$\Sigma_b(k)$ to denote the integral of $mn(m)\,b(m)\,W[kR(m)]$, then 
the remaining integral over $k$ is really of the form 
$\int{\rm d}k\ P(k)\,[\Sigma^2(k)+\Sigma_b^2(k)]\,K(kr)/2\pi^2$.  
The result of all this is that 
\begin{eqnarray}
S(m_1,m_2|r) &=& \sigma^2_{\rm vir}(m_1) + \sigma^2_{\rm vir}(m_2) \nonumber \\
&&\ + \sigma^2_{\rm halo}(m_1) + \sigma^2_{\rm halo}(m_2) - 2 \Psi(m_1,m_2|r) 
\nonumber \\ 
&&\ + {b(m_1)\,b(m_2)\ v^2_{12}(r)\over 2[1 + b(m_1)\,b(m_2)\xi_0(r)]},
\label{sm1m2r}
\end{eqnarray}
where the $\sigma$s are assumed to be the dispersions in one dimension.  
The final term comes from the fact that the statistic is not centred.  
It was obtained by expanding 
$\langle(v_1-v_2)^2(1+b_1\delta_1)(1+b_2\delta_2)\rangle$
in $\delta_1$ and $\delta_2$, and using the linear theory relation 
between $v$ and $\delta$ (e.g. Fisher 1995).  

\begin{figure}
\centering
\mbox{\psfig{figure=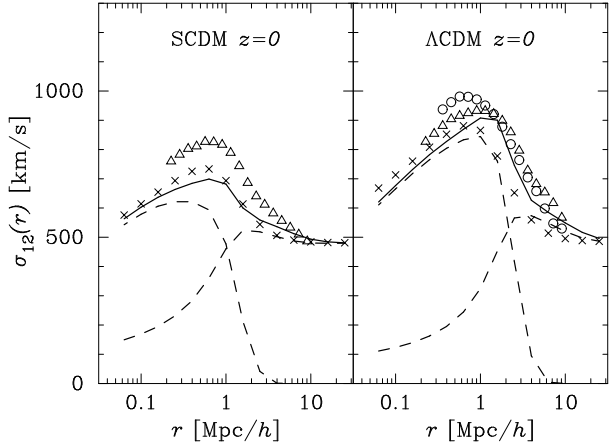,height=6cm,bbllx=72pt,bblly=58pt,bburx=629pt,bbury=459pt}}
\caption{Scale dependence of the pairwise velocity dispersion.  
Symbols, the same as in the previous figure, represent the simulation 
results.  Solid curves show the scale dependence in our model after 
including the fact that the dispersion within a halo depends on position 
within the halo, and also accounting crudely for the effects of spatial 
as well as velocity correlations.  
The two dashed curves show the contribution from pairs in which both 
particles are in the same halo (peak at small $r$) and in separate 
haloes (peak at large $r$).  }
\label{sig12c}
\end{figure}

Fig.~\ref{sig12c} compares our model with what is measured in simulations.  
The symbols are the same as in the previous figure; they are included to 
show the uncertainties associated with the measurements to date, and the 
influence of the finite size of the simulation box.  
As before, we have accounted for the finite box size (which we set to 
$L=141$ Mpc/$h$) when showing our model predictions.  
The two dashed curves are the model predictions for pairs in which both 
particles are in the same halo (peak at small $r$), and in different 
haloes (peak at large $r$).  The solid curve is the sum in quadrature 
of the two contributions, plus a piece which comes from the fact that 
the statistic is not centred.  

In addition to including the effects of velocity correlations (which 
affects the two-halo term), we have modified the one-halo term to include 
the fact the velocity dispersion within a halo depends on position within 
the halo, as described in the Appendix.  A glance at Fig.~\ref{jeans} shows 
that, for massive haloes, the isothermal assumption overestimates the 
dispersion.  Because the peak is mainly due to pairs in massive haloes, 
using what is, arguably, the more realistic value for the dispersion 
lowers the height of the peak.  With these two changes, our model falls 
significantly below the SCDM results.  In the $\Lambda$CDM model, on the 
other hand, our model appears to be in reasonable agreement with the 
simulations, though it too underestimates the simulation results, 
especially on small scales.  

\subsection{Dependence on trace-particle type}\label{types}
Our model has the virtue that it is straightforward to study how 
the pairwise velocity dispersion depends on the type of trace particle.  
For example, if only haloes were used to construct this statistic, one 
would expect the large scale asymptotic value of $\sigma_{12}$ to be 
smaller than for the dark matter, since, in this case, the virial term 
does not contribute to the dispersion.  
Figure~1 of Sheth \& Diaferio (2000) suggests that the term which 
remains depends slightly on halo mass, and that linear theory, smoothed 
on the appropriate scale, provides a reasonably good estimate of what 
it is.  

In addition to being smaller than the dark matter pairwise dispersion, 
the scale dependence of $\sigma_{12}(r)$ for haloes will also be 
different than it is for dark matter.  We can use the results above 
to estimate how it scales.  For halos, there is no contribution 
from virial motions, so 
\begin{eqnarray}
\sigma_{12}^{\rm halos}(r) &=& \int\!\!\int {\rm d}m_1\,{\rm d}m_2 \,
{1 + \xi_{\rm hh}(m_1,m_2|r)\over [1+\Xi_{\rm hh}(r)]} \nonumber \\
&&\qquad\times\ \ n(m_1)\,n(m_2)\,H(m_1,m_2|r)
\label{sig12h}
\end{eqnarray}
where 
\begin{displaymath}
1+\Xi_{\rm hh}\equiv \int\!\!\int {\rm d}m_1\,{\rm d}m_2 \, n(m_1)\,n(m_2)
\Bigl[1 + \xi_{\rm hh}(m_1,m_2|r)\Bigr],
\end{displaymath}
and $H(m_1,m_2|r)$ is given by equation~(\ref{sm1m2r}) with the 
virial terms set to zero.  A more exact expression, appropriate for peaks 
identified with the same smoothing scale ($m_1=m_2$ in our notation), has 
been provided by Reg\"os \& Szalay (1995):  our approximate expression is 
very similar to their equation~(68).

Equation~(\ref{sig12h}) shows that $\sigma_{12}^{\rm halos}$ at 
intermediate $r$ should be slightly smaller than the asymptotic large 
$r$ value, whereas the pairwise dispersion is larger at intermediate 
$r$ for the dark matter.  

Galaxies in massive haloes are essentially trace particles, so their 
motions are similar to the motions of dark matter particles 
(see Sheth \& Diaferio 2000 for some subtleties associated with 
motions of galaxies in less massive haloes).  
For the velocity dispersion statistic, the important difference between 
galaxies and dark matter particles is that, whereas the number of dark 
matter particles in a halo is proportional to halo mass, and the number 
of haloes in a halo is unity (of course!) for all haloes, the number of 
galaxies in a halo is something in between.  Essentially, this happens 
because the number of galaxies in a halo is proportional not to the 
total amount of gas in the halo, but to the amount of gas which can 
cool.  So the number of galaxies increases with halo mass, but not as 
quickly as the number of dark matter particles does.  
Thus, relative to the statistics of dark matter particles, galaxies 
downweight the contribution from massive haloes.  Since massive haloes 
have large virial motions, these are less pronounced for galaxies 
than for dark matter.  As a result, the pairwise dispersion of galaxies 
will be smaller in amplitude, and less scale-dependent, than that of 
the dark matter.  This was first noticed by Jing, Mo \& B\"orner (1998), 
who pointed out that this is what was required for consistency with 
observations.  

Peacock \& Smith (2000) used a simple $N_{\rm gal}(m)$ prescription 
to generate galaxies from their dark matter simulations.  They then 
measured the pairwise dispersion of the model galaxies in their 
simulations.  Our analytic estimate of the pairwise dispersion of 
the dark matter allows us to do analytically what Peacock \& Smith 
did numerically.  First, we use the $N_{\rm gal}(m)$ relation to compute 
the correlation function of galaxies.  Essentially, this can be done by 
setting $\rho(r|m) \to \rho(r|m)\,N_{\rm gal}(m)/m$ in 
Section~\ref{modelxi} (but see Seljak 2000, Peacock \& Smith 2000, or 
Scoccimarro et al. 2000 for some subtleties associated with how exactly 
this is done).  We then use this pair-weighting to compute the velocity 
dispersion of our model galaxies.  

Specifically, we set 
\begin{eqnarray}
\sigma_{\rm g1halo}^2(r) &=& {1 \over {1+\xi_{\rm gal}(r)}}
\int {\rm d}m \, {Gm\over r_{\rm vir}} \\ \nonumber 
&& \times {\langle N_{\rm gal}^2(m)\rangle\over m^2}
{n(m)\over\bar n_{\rm gal}}\,{\lambda(r|m)\over\bar n_{\rm gal}}
\end{eqnarray}
where $\bar n_{\rm gal} \equiv \int {\rm d}m \, m n(m) N_{\rm gal}(m)/m$ 
and $\xi_{\rm gal}(r)$ is the galaxy correlation function 
(see e.g. Seljak 2000; Scoccimarro et al. 2000), and 
\begin{eqnarray}
\sigma^2_{\rm g2halo}(r) &=& \int\!\!\int {\rm d}m_1 {\rm d}m_2
{1 + \xi(m_1,m_2|r)  \over [1 + \xi_{\rm gal}(r)]} \nonumber \\ 
&& \times \left [ N_{\rm gal}(m_1) \over m_1 \right] 
\left [ N_{\rm gal}(m_2) \over m_2 \right]\nonumber\\
&&\ \ \times \ \ 
{m_1 n(m_1)\over\bar n_{\rm gal}}\,
{m_2 n(m_2)\over\bar n_{\rm gal}}\,S(m_1,m_2|r) ,
\end{eqnarray}
where $S(m_1,m_2|r)$ is as in equation (\ref{sm1m2r}).

\begin{figure}
\centering
\mbox{\psfig{figure=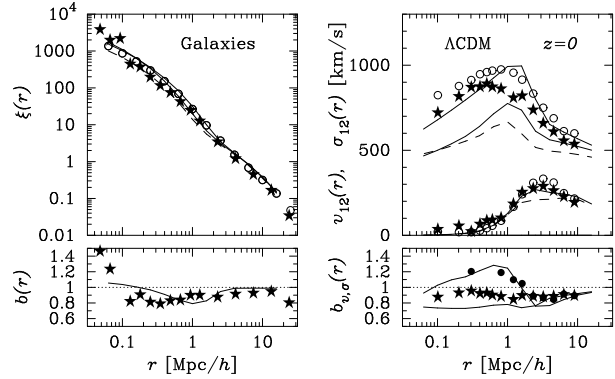,height=5cm,bbllx=69pt,bblly=58pt,bburx=721pt,bbury=465pt}}
\caption{The correlation function and pairwise velocity dispersion 
for simple models of galaxies.  Open circles in both panels show 
the statistics of the dark matter in the $\Lambda$CDM GIF simulation, 
and stars show the corresponding statistics computed using the semianalytic 
bright, extinction corrected galaxies of Kauffmann et al. (1999), measured 
in the same simulation.  Solid curves show our model predictions.  
The two bottom panels show the square root of the ratio of the galaxy 
and dark matter correlation functions, and the ratio of the streaming 
motions and rms pairwise velocities, respectively.  Dashed curves in the 
top panels show our model predictions for the semianalytic models of 
Benson et al. (2000).}
\label{gals}
\end{figure}

To illustrate, Fig.~\ref{gals} shows the result of doing this using 
the $N_{\rm gal}(m)$ relation obtained from the publically available 
GIF $\Lambda$CDM semi-analytic galaxy formation models of 
Kauffmann et al. (1999).  We actually used the simple fit to this 
relation, for galaxies brighter than $M_B\le -17.5 + 5\log h$ after 
correcting for the effects of dust, provided by Sheth \& Diaferio (2000).  
The panel on the left shows the correlation functions of the dark matter 
(open circles) and the galaxies (stars), and our models for the two 
correlation functions (solid lines).  The bottom left panel shows the 
square root of the ratio of the two correlation functions.  

The open circles in the panel on the right show the mean streaming 
motions and the pairwise dispersions of the dark matter particles, 
and the stars show the corresponding statistics for the semianalytic 
galaxies.  We assigned velocities to the galaxies slightly differently 
than how Kauffmann et al. (1999) did.  Namely, for the central galaxy 
in a halo, and for all haloes in which there was only one galaxy, we used 
the halo centre of mass velocity, rather than the velocity of the nearest 
particle to represent the motion of the galaxy.  The difference between 
the two speeds is typically about 80 km/s; this means that our values 
for the mean streaming and velocity dispersion statistics are slightly 
smaller than the ones presented in Fig.~13 of Kauffmann et al. (1999).  

The bottom right panel shows the ratio of the mean streaming motions of 
galaxies to that of the dark matter (filled circles) and the ratio of 
the rms pairwise velocity of galaxies to that of the dark matter (filled 
stars).  Solid lines show our model predictions.  We used the simpler 
isothermal approximation, rather than the actual position dependent 
dispersion, when computing the model predictions.  Both in the simulations 
and in our model, the pairwise dispersion for the galaxies falls below 
that of the dark matter, although our model overestimates the amount 
by which this happens.  The streaming motions of galaxies are similar 
to those of the dark matter on large scales because the galaxies have 
a bias factor which is close to unity.  Presumably, if the large scale 
bias factor were different from unity, the streaming motions on large 
scales would also be different from the dark matter.  On small scales, 
however, the galaxy mean streaming motions can be rather different from 
that of the dark matter; our model is able to provide quite a good 
description of this difference.  

Benson et al. (2000) showed that the pairwise dispersions of 
galaxies in their semianalytic galaxy formation models are lower 
than those of Kauffmann et al. (1999), and that the reason for this 
was because the two models have different $N_{\rm gal}(m)$ relations.  
Therefore, we also looked at a model in which 
 $N_{\rm gal}(m)=(m/10^{12.9} M_\odot/h)^{0.6}$ 
for haloes more massive than $3\times 10^{11} M_\odot/h$; 
this provides a good fit to the Benson et al. (2000) models with 
$M_B\le -19.5 + 5\log h$, and is similar to the sort of scaling which 
Jing, Mo \& B\"orner (1998) argued was required for agreement with 
observations.  The dashed curves in the two upper panels of 
Fig.~\ref{gals} show our model predictions when this relation is used.  
The correlation functions of the two semi-analytic models are similar 
despite the different brightness cuts.  However, because this model has 
fewer galaxies in massive haloes than the Kauffmann et al. model, the 
pairwise dispersion of the galaxies in this model is lower.  This 
is because, compared to the correlation function, the pairwise dispersion 
has an extra $Gm/r_{\rm vir}\propto m^{2/3}$ dependence on halo mass, so 
it is that much more sensitive to the presence or absence of massive haloes.  
The Benson et al. models for galaxies with $M_B\le -18.5 + 5\log h$ 
are well fit by  $N_{\rm gal}(m)=(m/10^{12.6} M_\odot/h)^{0.75}$.  
If we use this relation instead, then the amplitude of $\sigma_{12}$ 
increases.  This is consistent with what Benson et al. measured 
in their simulations.  

\section{Discussion}
Writing the correlation function as the sum of two terms, one which is 
essentially described by linear theory, and dominates on large scales, 
and another which is inherently nonlinear, and dominates on small scales, 
is both accurate and useful.  Such a split has been used to model the 
spatial distribution of the dark matter and of galaxies (Seljak 2000; 
Peacock \& Smith 2000; Scoccimarro et al. 2000).  Sheth \& Diaferio (2000) 
show that a similar split can be applied to velocities; particle velocities 
can also be written as the sum of linear and nonlinear parts.  In this 
paper, we used this split to compute simple estimates not just of the 
correlation function of dark matter particles, but of the single particle 
velocity dispersion (the density weighted temperature) and its evolution, 
and the scale dependence of the first and second moments of the pairwise 
velocity distribution as well.  

Our model provides a good description of the scale dependence of 
the mean streaming velocities (Figs.~\ref{v12plot} and~\ref{v12Hr}).  
It also provides a reasonably good description of the scale dependence 
of the pairwise dispersion, provided one restricts attention to scales 
larger than a Megaparsec or so (Figs.~\ref{sig12} and~\ref{sig12c}); 
our model appears to underestimate the value of $\sigma_{12}$ on scales 
smaller than 1 Mpc$/h$.  This is, perhaps, surprising, because on these 
smaller scales, most pairs are in the same halo, and one might have 
thought that virialized haloes were rather simple to model.  
Our model assumes that virialized haloes are smooth 
(or, more importantly, that substructure does not substantially affect 
the velocity dispersion), and that the velocity dispersion within them 
is isotropic.  The discrepancy between our model and the simulations 
suggests that one or both of these assumptions is wrong.  To model the 
small separation regime accurately, we may have to resort to using the 
exact approach based on the BBGKY hierarchy (e.g. Peebles 1980).
As we mentioned in the introduction, this approach is considerably more 
complicated, because it requires knowledge of the three-point correlation 
function.  Although this can be done within the context of these halo 
models using results presented in Scoccimarro et al. (2000), we have not 
done so here.  

Despite the shortcomings of our model for the pairwise dispersion 
on small scales, we feel that there are at least two reasons why 
it is useful.  First, it is much simpler than the exact approach one 
is led to from the BBGKY hierarchy, for which knowledge of the 
three-point correlation function is required.  Secondly, it is easily 
extended to provide predictions as a function of trace particle type.  
This is particularly useful for comparing theoretical models with 
observations of galaxies.  Section~\ref{types} showed how to model 
the dependence on separation of the galaxy pairwise velocity dispersion.  

The main reason for doing this was the following.
Our model for the linear and nonlinear contributions to the number of 
pairs at a given separation can be combined with Sheth \& Diaferio's 
model for the linear and nonlinear contributions to velocities, to model 
how the full distribution of pairwise velocities (not just the first and 
second moments) depend on pair separation.  Such a model can be 
extended to describe galaxies, just as we did here.  
Following Fisher (1995), this allows us to estimate the effect of 
redshift space distortions on the shape of the galaxy correlation 
function over the entire range of linear to nonlinear scales.  
This is the subject of work in progress.  

\section*{Acknowledgments}
This collaboration was started at the German American Young Scholars
Institute in Astroparticle Physics, at the Aspen Center for Physics
in the fall of 1998, and at Ringberg Castle in 1999.
RKS thanks the IAS, as well as the University and Observatory at Torino, 
and LH \& RS thank Fermilab for hospitality where parts of this work 
were done.  We all thank the Halo Pub for inspiring nourishments.
The N-body simulations, halo and galaxy catalogues used in this paper
are publically available at {\tt http://www.mpa-garching.mpg.de/NumCos}.
The simulations were carried out at the Computer Center of the 
Max-Planck Society in  Garching and at the EPCC in Edinburgh, as part 
of the Virgo Consortium project.  In addition, we would like to thank 
the Max-Planck Institut f\"ur Astrophysik where some of the computing 
for this work was done.  
RKS is supported by the DOE and NASA grant NAG 5-7092 at Fermilab. 
LH is supported by NASA grant NAG5-7047, NSF grant PHY-9513835
and the Taplin Fellowship.  He also thanks the DOE for an 
Outstanding Junior Investigator Award (grant DE-FG02-92-ER40699). 
RS is supported by endowment funds from the IAS.

\appendix
\section{Halo profiles}
We study a number of different halo profiles below.  
For each profile, we will be interested in how the density $\rho$, 
the circular velocity $v_{\rm c}$, the radial velocity dispersion 
$\sigma^2_r$, and the potential $\phi$ depend on distance from the 
halo centre.  In addition, we will be interested in the convolution 
of a profile with itself: $\lambda$.  

The circular velocity at a distance $r$ from the centre of a halo is 
defined as the square root of the ratio of the mass interior to $r$ 
and $r$:  
\begin{equation}
v_{\rm c}^2(r|m) \equiv {G\,m(<r)\over r} 
= {4\pi G\over r}\int_0^r {\rm d}x\ x^2\,\rho(x|m).
\label{vcr}
\end{equation}
This, averaged over the profile shape, is 
\begin{equation}
mV_{\rm c}^2(m) \equiv 
4\pi\ \int {\rm d}r\ r^2\,\rho(r|m)\,v_{\rm c}^2(r|m) .
\end{equation}
The potential energy at $r$ is 
\begin{eqnarray}
\phi(r|m) &=& {-4\pi G\over r}\int_0^r {\rm d}r'\ {r'}^2\,\rho(r'|m) 
\nonumber\\
&&\qquad \ - 4\pi G\int_r^\infty {\rm d}r'\ r'\,\rho(r'|m).
\end{eqnarray}
The total potential energy is this, integrated over the halo: 
\begin{equation}
W(m) = 2\pi \int {\rm d}r\ r^2\,\rho(r|m)\,\phi(r|m) = -mV_{\rm c}^2(m),
\end{equation}
where the final equality follows after rearranging the order of the 
integrals.  
If a halo is assumed to be in equilibrium, and the orbits within it 
are assumed to be isotropic, then the Jeans equation can be used to 
compute the radial velocity dispersion $\sigma_r^2$:
\begin{equation}
-{\rm d\,\rho\sigma_r^2\over {\rm d}r} = 
\rho(r|m)\,{\rm d\,\phi\over {\rm d}r} = \rho(r|m)\,{Gm(<r)\over r^2}.
\end{equation}  
The total kinetic energy of the halo is 
\begin{equation}
K(m) = {3\over 2}\ 4\pi \int {\rm d}r\ r^2\,\rho(r|m)\sigma_r^2(r|m) 
     = {3\over 2}\,{mV^2_{\rm c}(m)\over 3}.
\end{equation}
These relations between $W$, $K$ and the circular velocity are true 
for all profiles, and will be a useful consistency check in what 
follows.  Notice that $-W(m) = 2K(m)$: the haloes are in virial 
equilibrium.  

\subsection{The Hernquist profile}\label{hquist}
The mass associated with this profile is finite even though 
the halo extends smoothly to infinity.  For this reason, it will 
be a very useful benchmark calculation in what follows.  
For a halo of mass $m$ at $r_{\rm vir}$, the profile is 
\begin{equation}
{\rho_{\rm H}(s)\over\bar\rho} = 
{\Delta_{\rm vir}\over 3\Omega} {2b\,(1+b)^2\over s\,(b + s)^3} = 
{\Delta_{\rm vir}\over 3\Omega} {2\,(1+b)^2\over b^3\,x\,(1 + x)^3} ,
\label{rhoh}
\end{equation}
where $\Delta_{\rm vir}$ is the average density within $r_{\rm vir}$ 
in units of the critical density, $s=r/r_{\rm vir}$, $b$ is a core 
radius in units of $r_{\rm vir}$, and $x = s/b$.  
The mass interior to $r$ is given by 
\begin{equation}
{m(<r)\over m} = {(1+b)^2\,x^2\over (1+x)^2},
\end{equation}
the circular velocity is 
\begin{equation}
v_c^2(r) = {Gm(<r)\over r} 
         = {Gm\over r_{\rm vir}}\,{x\,(1+b)^2\over b\,(1+x)^2},  
\end{equation}
and the radial velocity dispersion $\sigma_r^2$, which can be computed 
from the Jeans equation, is also analytic:
\begin{eqnarray}
{\sigma_r^2(r)\over Gm/r_{\rm vir}} &=& {(1+b)^2\over 12 b}
\Bigl[12 x(1+x)^3\,{\rm ln}\left({1+x\over x}\right) - \nonumber\\
&&{25x + 52 x^2 + 42 x^3 + 12x^4\over 1+x}\Bigr]
\end{eqnarray}
(Hernquist 1990; Cole \& Lacey 1996).  
The potential energy at $r$ of such a halo is 
\begin{equation}
\phi(r) = -{Gm\over r_{\rm vir}}\,{(1+b)^2\over b\,(1+x)} ,
\end{equation}
and so the total potential energy of the halo is 
\begin{equation}
W_{\rm H} = {4\pi\over 2}\int {\rm d}r\ r^2\,\rho_{\rm H}(r)\,\phi(r) 
          = -{G m^2\over 6r_{\rm vir}} {(1 + b)^4\over b} 
\label{whern}
\end{equation}
(Hernquist 1990).  It is straightforward to verify that 
the kinetic energy $K_{\rm H} = -W_{\rm H}/2$:  the halo is in 
virial equilibrium.  A similar averaging of $v_c^2(r)$ equals 
$-W_{\rm H}$.  

The convolution of such a profile with itself is 
\begin{equation}
{4\pi r_{\rm vir}^3\,b^3\over m^2} {\lambda_{\rm H}(r|m,b)\over (1+b)^4} 
 = {4\over x^4}\ {h_1(x) - h_2(x)\over (2+x)^4},
\end{equation}
where
\begin{eqnarray*}
h_1(x) &=& {24 + 60 x + 56 x^2 + 24 x^3 + 6 x^4 + x^5\over 1 + x} \ \ 
{\rm and}\nonumber\\
h_2(x) &=& {12\,(1 + x)\,(2 + 2 x + x^2)\,{\rm ln}(1 + x)\over x}.
\end{eqnarray*}
A little algebra shows that 
$2\pi G \int {\rm d}r\ r\,\lambda_{\rm H}(r|m,b)$ equals 
the potential energy of the halo, as it should.  

\subsection{The truncated singular isothermal sphere}\label{isotmal}
In this case a halo of mass $m$ is truncated at its virial 
radius $r_{\rm vir}$.  On scales smaller than this, 
\begin{equation}
{\rho(r)\over\bar\rho} = {\Delta_{\rm vir}/3\Omega\over s^2},
\end{equation}
where $s=r/r_{\rm vir}$, and $\Delta_{\rm vir}$ is a constant 
which specifies how dense the halo is relative to the critical 
density at the time:  
 $3 m/4\pi r^3_{\rm vir} = \Delta_{\rm vir}\rho_{\rm crit}$.  
The circular velocity within the halo is 
\begin{equation}
v_{\rm c}^2 = Gm/r_{\rm vir}, 
\end{equation}
and the radial velocity dispersion within the halo is 
$\sigma_r^2 = v_{\rm c}^2/2$; for an isothermal sphere, $v_{\rm c}$ 
and $\sigma_r$ are independent of position 
within the halo.  The convolution of such a profile with itself is 
\begin{equation}
{4\pi r_{\rm vir}^3\over m^2}\,\lambda_{\rm Iso}(r|m) = {1\over s}\,
\int_{s/2}^{1} {{\rm d}x\over x}\,{\rm ln}\Bigl|{x\over s-x}\Bigr|
\ \ {\rm if}\ 0\le s\le 2,
\end{equation}
and the truncation means that it is zero on separations larger 
than twice $r_{\rm vir}$.  So, for a truncated isothermal sphere 
\begin{equation}
2\pi G \int {\rm d}r\ r\,\lambda_{\rm Iso}(r|m,b) 
= {G m^2\over r_{\rm vir}};
\end{equation}
this equals the circular velocity averaged over the halo, 
which in turn equals $2K_{\rm Iso}=-W_{\rm Iso}$.  

\subsection{The truncated NFW profile}\label{nfw}
We could go through a similar exercise for a truncated Hernquist 
profile.  Instead, we will study another profile which declines 
slightly less steeply at the edge, and so is able to fit the results 
of numerical simulations slightly better (Navarro, Frenk \& White 1997).  
The NFW profile contains mass $m$ within $r_{\rm vir}$, and it is 
truncated at this scale.  Within the virial radius, 
\begin{equation}
{\rho(r)\over\bar\rho} = {\Delta_{\rm vir}\over 3\Omega}
{f(a)\over a^3 x\,(1 + x)^2}
\end{equation}
where $x = r/a$, with $s=r/r_{\rm vir}$ as before, 
$a$ is the core radius in units of $r_{\rm vir}$, and 
\begin{equation}
f(a) = \Bigl[{\rm ln}(1 + 1/a) - 1/(1+a)\Bigr]^{-1}.  
\end{equation}
The potential at $r$ is 
\begin{equation}
\phi(r) = -{Gm\over r_{\rm vir}}\,{f(a)\over a}\,
\left[{{\rm ln}(1+x)\over x} - {a\over 1+a}\right] .
\label{phitrunc}
\end{equation}
The corresponding expression in Cole \& Lacey (1996) does not have 
the second term in the square brackets because we are assuming the halo 
is truncated at the virial radius, whereas they did not.  The circular 
velocity is 
\begin{equation}
v_{\rm c}^2(r) = {Gm\over r_{\rm vir}}\,{f(a)\over a}\, 
\left[{{\rm ln}(1 + x)\over x} - {1\over 1 + x} \right],
\end{equation}
and the total potential energy is 
\begin{equation}
W_{\rm NFW} = -{G m^2\over 2\, r_{\rm vir}}\
   {f(a)^2\over a(1+a)}\left[1 - 2a\,{\rm ln}(1 + 1/a) + {a\over 1+a}\right].
\label{wnfw}
\end{equation}
for the truncated potential given above.  
This equals the average of the circular velocity over the halo.
If the halo is not truncated (equation~\ref{phitrunc} without the second 
term in the square brackets), then the total potential energy is 
$W_{\rm NFW}=- (G m^2/r_{\rm vir})\,f^2(a)/2$ if we integrate over all 
$r$, and it is 
\begin{equation}
W_{\rm NFW} = -{G m^2\over 2\, r_{\rm vir}}\
   {f(a)^2\over a(1+a)}\left[1 - a\,{\rm ln}(1 + 1/a) \right],
\end{equation}
if we integrate out to $r_{\rm vir}$ only.

\begin{figure*}
\centering
\mbox{\psfig{figure=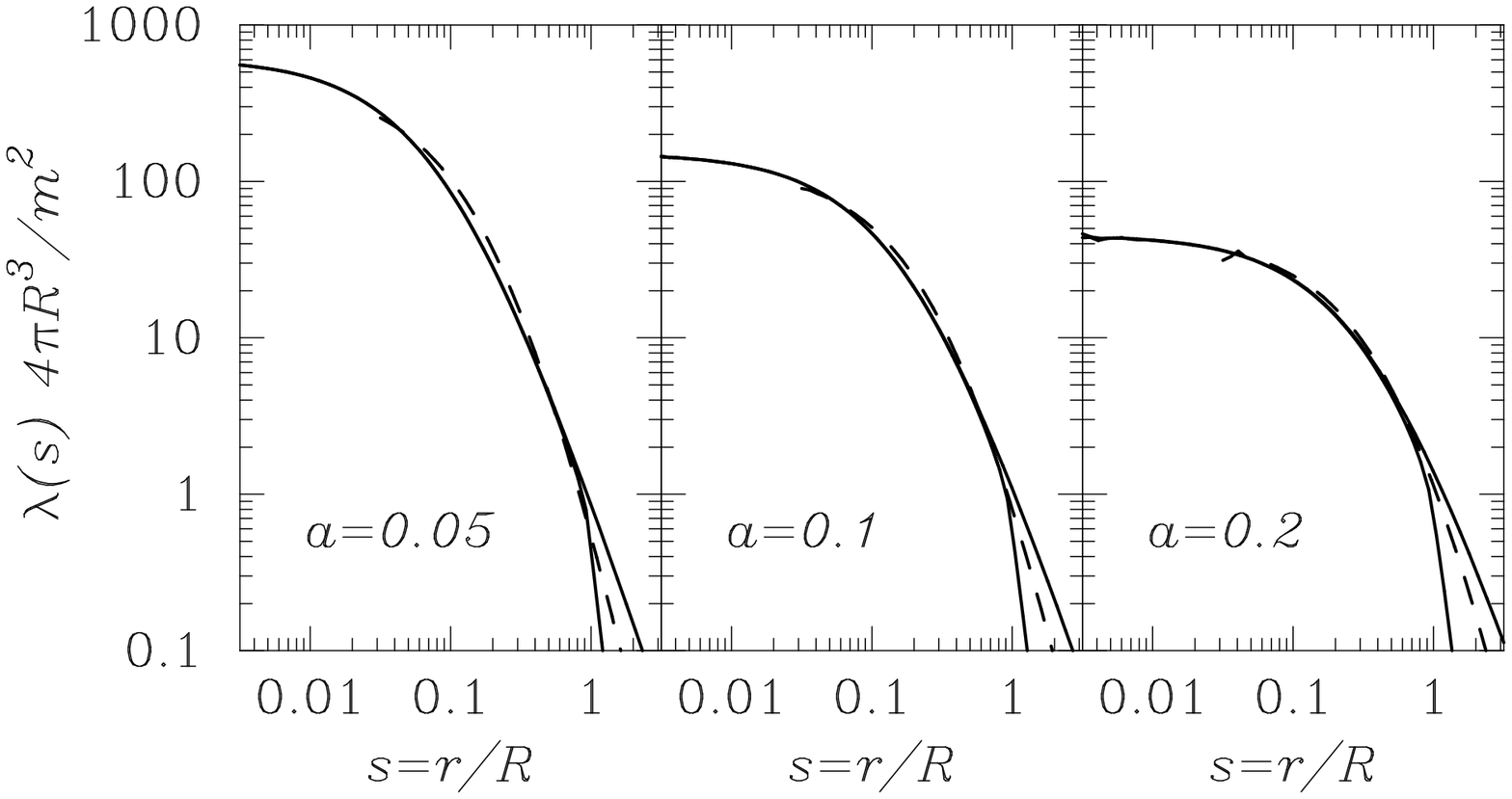,height=7cm,bbllx=63pt,bblly=115pt,bburx=612pt,bbury=396pt}}
\caption{Shape of the convolution of the density profile as a function 
of distance in units of the virial radius for a few representative 
values of the core radius $a$.  
Small values of $a$ (left panel) correspond to low mass haloes, whereas 
more massive haloes have larger core radii (right panel).  Solid curves 
show the result for NFW haloes truncated at the virial radius (lower 
amplitude at large $r$) and when the haloes are allowed to extend to 
infinity (larger amplitude at large $r$), and dashed curves show the 
corresponding result for a Hernquist profile which contains the same 
mass within the virial radius.  The core radius of the Hernquist profile 
is $\sqrt{2}\,a^{0.75}$.}
\label{lambdas}
\end{figure*}
\begin{figure*}
\centering
\mbox{\psfig{figure=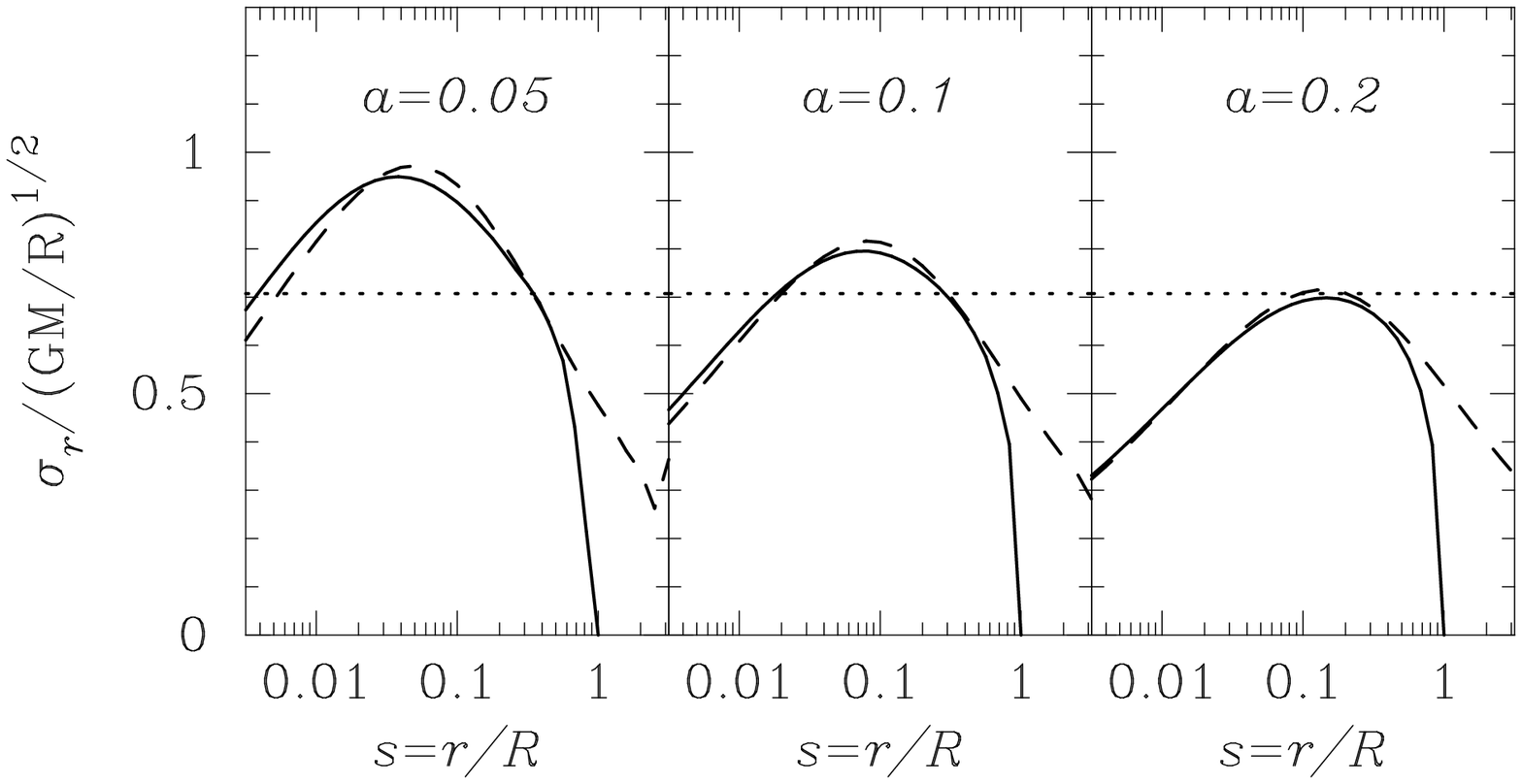,height=7cm,bbllx=63pt,bblly=115pt,bburx=612pt,bbury=396pt}}
\caption{Radial velocity dispersion as a function of distance from 
halo centre if orbits are isotropic for the same models as the previous 
figure.  Solid curves show the dispersion in NFW haloes truncated at the 
virial radius, dotted curves show the isothermal value (which is 
independent of distance from centre) and dashed curves show the 
result for the corresponding Hernquist profile. }
\label{jeans}
\end{figure*}

The radial velocity dispersion, computed from the Jeans equation, is
\begin{equation}
{\sigma_r^2(r)\over Gm/r_{\rm vir}} = {f(a)^2\over 2a}\ 
{x(1+x)^2\over f(a)} \ \Bigl[ g(1/a)-g(x) \Bigr],
\label{nfwjeans}
\end{equation}
where  
\begin{eqnarray*}
g(x) &\equiv& -1 + \frac{1}{x} +
\frac{1}{(1+x)^2}+\frac{6}{1+x}+\ln \frac{x}{1+x} \nonumber\\
&&\ + \frac{6x^2+3x-1}{x^2(1+x)} \ln(1+x) -3 \ln^2 (1+x) + 6 {\rm Li}_2(x):
\end{eqnarray*}
the dilogarithm ${\rm Li}_2(x)$ is defined by
\begin{displaymath}
{\rm Li}_2(x) \equiv \int_0^x {\rm d}\ln\,z\ \ln(1+z).
\end{displaymath}
This, in the expression for the total kinetic energy shows 
explicitly that $-W_{\rm NFW} = 2K_{\rm NFW}$.  This relation 
is satisfied exactly because we truncated the halo at the virial 
radius.  If we do not truncate, and we let the halo extend to 
infinity, then the expression above should be modified by setting 
$g(1/a)\to \pi^2 - 1$.  

The convolution of such a profile with itself gives 
\begin{eqnarray}
{4\pi r_{\rm vir}^3a^3\over m^2}\,{\lambda_{\rm NFW}(r|m,a) \over f(a)^2}
&=& {-4(1+a) + 2ax(1+2a) + a^2x^2\over 2x^2 (1 + a)^2 (2 + x)}
\nonumber\\
&& + {1\over x^3}\,{\rm ln}\left[{(1+a-ax)(1+x)\over (1+a)}\right]\nonumber \\
& & + {{\rm ln}(1+x)\over x (2+x)^2} \ \ {\rm if}\ 0\le s\le 1 
\nonumber
\end{eqnarray}
\begin{eqnarray}
\ &=& {{\rm ln}[(1+a)/(ax+a-1)]\over x(2+x)^2 }  
+ {a^2x-2a \over 2 x (1+a)^2 (2+x)} \nonumber \\
&&\qquad\qquad \ {\rm if}\ 1\le s\le 2.
\label{lnfw}
\end{eqnarray}
In the limit in which the virial radius is much larger than the core 
radius, this separates into the product of a function of $a$ and 
another of $x$:  
\begin{eqnarray*}
\lambda_{\rm NFW}(r|m,a)&\to& {m^2\,f(a)^2\over 4\pi r_{\rm vir}^3a^3}\,
{2\over x^2(x+2)}\nonumber\\
&&\ \times\ \left[{(x^2 + 2x+2)\ln(1+x)\over x(x+2)}-1\right].
\end{eqnarray*}
Straightforward but tedious algebra shows that the integral of 
$2\pi G r^2\,\lambda_{\rm NFW}(r|m,a)/r$ equals $W_{\rm NFW}$, 
the average of the potential over the halo.  Again, this equality 
is exactly satisfied only because we have self-consistently 
truncated our haloes (equations~\ref{lnfw} and~\ref{wnfw}).  

It is an interesting question as to whether or not truncation is 
important.  Although they discuss NFW haloes truncated at the virial 
radius (so they have finite mass) it appears that Cole \& Lacey (1996) 
do not truncate their NFW haloes when computing $\phi$ and $\sigma_r$.  
As a result, their values of $\sigma_r$ do not go to zero at the virial 
radius, and their values of $W_{\rm NFW}/2K_{\rm NFW}$ at the virial 
radius are slightly greater than unity.  Both these are in agreement 
with the simulation results they present.  Since NFW haloes are not
really isolated objects, it may be that Cole \& Lacey's decision 
to ignore the truncation at the virial radius when computing all 
quantities except the mass is more physically reasonable.  

Deciding whether to truncate or not is important if one wishes to 
include the scale dependence of the velocity dispersion within a halo 
into our model for the pairwise velocity dispersion.  We found that 
using our formula for truncated NFW haloes, and then integrating over
the distribution of halo masses produced values of $\sigma_{12}$ which 
were about 15\% lower than the simulation results presented in 
Fig.~\ref{sig12c}.  If we do not truncate, our models are better able 
to reproduce the measurements in simulations, suggesting that this is, 
indeed, the correct thing to do.  

The solid curves in Fig.~\ref{jeans} show the radial velocity dispersion 
as a function of radius for truncated NFW haloes (eq.~\ref{nfwjeans}) 
for three representative values of the concentration parameter $a$;
panels on the left correspond to the least massive haloes.  
The dotted curves show the isothermal value:  one half of 
$GM/r_{\rm vir}$, and dashed curves show the corresponding result 
for an infinite Hernquist profile, with core radius 
$b = \sqrt{2}\,a^{0.75}$.  With this scaling, the Hernquist formula
for the dispersion provides a good approximation to what happens if 
we remove the condition that the NFW profile is truncated.  This is 
useful because the Hernquist radial velocity profile is analytic.  


\begin{thebibliography}{99}
\bibitem{durham} Benson A. J., Baugh C. M., Cole S., Frenk C. S., 
Lacey C. G., 2000, MNRAS, 311, 793
\bibitem{dmw} Davis M., Miller A., White S. D. M., 1997, ApJ, 490, 63
\bibitem{dg96} Diaferio A., Geller M., 1996, ApJ, 467, 19
\bibitem{karl} Fisher K. B., 1995, ApJ, 448, 494
\bibitem{kris} G\'orski K., 1988, ApJ, 332, L7
\bibitem{hklm} Hamilton A. J. S., Kumar P., Lu E., Matthews A., 1991, ApJ, 
374, L1
\bibitem{virgo} Jenkins A., Frenk C. S., Pearce F. R., Thomas P. A., Colberg
J. M., White S. D. M., Couchman H. M. P., Peacock J., A., Efstathiou G., 
Nelson A. H., 1998, ApJ, 499, 20
\bibitem{jmb} Jing Y. P., Mo H. J., B\"orner G., 1998, ApJ, 494, 1
\bibitem{jfs} Juszkiewicz R., Fisher K. B., Szapudi I., 1998, ApJ, 504, L1
\bibitem{jsd} Juszkiewicz R., Springel V., Durrer R., 1999, ApJ, 518, L25
\bibitem{gif} Kauffmann G., Colberg J. M., Diaferio A., White S. D. M., 
1999, MNRAS, 303, 188
\bibitem{mf} Ma C. P., Fry J., 2000, ApJ, submitted
\bibitem{mcs77} McClelland J., Silk J.,  1977, ApJ, 217, 331
\bibitem{mw96} Mo H. J., White S. D. M., 1996, MNRAS, 282, 347
\bibitem{mjb} Mo H. J., Jing Y. P., B\"orner G., 1997, MNRAS, 286, 979
\bibitem{ns59} Neyman J., Scott, E. L., 1959, Handbuch Phys., 53, 416
\bibitem{np94} Nityananda R., Padmanabhan T.,  1994, MNRAS, 271, 976
\bibitem{nfw95} Navarro J., Frenk C., White S. D. M., 1997, ApJ, 490, 493
\bibitem{paddy} Padmanabhan T., 1996, MNRAS, 278, 29P
\bibitem{jp} Peacock J., Smith R., 2000, MNRAS, submitted, astro-ph/0005010
\bibitem{peeb} Peebles P. J. E.,  1980, The Large Scale Structure of the 
Universe. Princeton Univ. Press, Princeton
\bibitem{ps74} Press W., Schechter P., 1974, ApJ, 187, 425
\bibitem{rs95}  Reg\"os E., Szalay A. S., 1995, MNRAS, 272, 447
\bibitem{uros} Seljak U., 2000, MNRAS, in press, astro-ph/0001493
\bibitem{halo} Scoccimarro R., Sheth R. K., Hui L., Jain B., 2000, ApJ,
accepted, astro-ph/0006319
\bibitem{rs96} Sheth R. K., 1996, MNRAS, 279, 1310
\bibitem{sj97} Sheth R. K., Jain B., 1997, MNRAS, 285, 231
\bibitem{sl98} Sheth R. K., Lemson G., 1999, MNRAS, 304, 767
\bibitem{st99} Sheth R. K., Tormen G., 1999, MNRAS, 308, 119
\bibitem{sd00} Sheth R. K., Diaferio A., 2000, MNRAS, submitted
\end{thebibliography}
\end{document}